\documentclass[fleqn,usenatbib]{mnras}
\usepackage{adjustbox}

\usepackage{adjustbox}
\usepackage{hyperref}
\hypersetup{colorlinks=true,linkcolor=magenta,filecolor=blue, urlcolor=magenta}
\usepackage{amsmath}
\usepackage{multirow}
\usepackage{booktabs}
\usepackage{natbib}
\numberwithin{equation}{section}
\def\setsymbol#1#2{\expandafter\def\csname #1\endcsname{#2}}
\def\getsymbol#1{\csname #1\endcsname}

\newbox\tablebox    \newdimen\tablewidth
\def\leaderfil{\leaders\hbox to 5pt{\hss.\hss}\hfil}

\def\tablenote#1 #2\par{\begingroup \parindent=0.8em
    \abovedisplayshortskip=0pt\belowdisplayshortskip=0pt
    \noindent
    $$\hss\vbox{\hsize\tablewidth \hangindent=\parindent \hangafter=1 \noindent
    \hbox to \parindent{$^#1$\hss}\strut#2\strut\par}\hss$$
    \endgroup}

\usepackage{newtxtext,newtxmath}

\usepackage[T1]{fontenc}

\DeclareRobustCommand{\VAN}[3]{#2}
\let\VANthebibliography\thebibliography
\def\thebibliography{\DeclareRobustCommand{\VAN}[3]{##3}\VANthebibliography}

\usepackage{graphicx}	

	\title[Dynamics of Gaussian Dark Energy EoS]{Probing the Dynamics of Gaussian Dark Energy Equation of State Using DESI BAO}
	
	\author[Hussain et al.]{Saddam Hussain$^1$, Simran Arora$^2$\thanks{\href{mailto:arora.simran@yukawa.kyoto-u.ac.jp}{arora.simran@yukawa.kyoto-u.ac.jp}}, Anzhong Wang$^3$, Benjamin Rose$^3$
    \\ 	
	$^1$Institute for Theoretical Physics and Cosmology, Zhejiang University of Technology, Hangzhou 310023, China.\\
	$^2$Center for Gravitational Physics and Quantum Information, Yukawa Institute for Theoretical Physics, Kyoto University, 606-8502, Kyoto, Japan.\\
$^{3}$ GCAP-CASPER, Department of Physics and Astronomy, Baylor University, Waco, TX 76798-7316, USA.}

\begin{document}
\label{firstpage}
\pagerange{\pageref{firstpage}--\pageref{lastpage}}
\maketitle

	\begin{abstract}
		We present an updated reconstruction of the DE equation of state (EoS), $w(a)$, employing the newly released DESI DR2 Baryon Acoustic Oscillation data. This analysis constrains the cosmological scenarios influenced by different models through the joint examination of a range of recently available cosmological probes, specifically the Pantheon+ sample and the DESY5 sample of Type Ia Supernovae, baryon acoustic oscillations, Hubble parameter measurements derived from cosmic chronometers, and cosmic microwave background distance priors based on the Planck 2018 data. Furthermore, we provide a concise perspective on the dynamical evolution of all models (CPL, PADE, GEDE, GDE, BellDE) and their interrelations. A Bayesian inference procedure is adopted to estimate the models parameters that yield the best fit to the data. The EoS remains within the phantom regime at higher redshifts, while favoring the quintessence regime in the current epoch. In this context, we propose a new Gaussian-like form of EoS, termed BellDE, which avoids phantom behavior (\(w \geq -1\)) at higher redshifts while remaining precisely calibrated at lower redshifts. Interestingly, BellDE exhibits a transient phantom nature (\(w < -1\)) around the transition redshift \(z \sim 0.5\), subsequently evolving into a quintessential regime (\(w > -1\)). In particular, the BellDE model provides competitive statistical preference while offering greater flexibility in the redshift regime $z \sim 0.5-1$, where DE is observationally significant.
\end{abstract}	

	
	
	

	

	\maketitle
	
	
	
	\section{Introduction}
	\label{introduction}

	Over the past two decades, a wealth of observational evidence, from Type Ia supernovae (SNeIa) \citep{SupernovaSearchTeam:1998fmf,SupernovaCosmologyProject:1998vns,Planck:2018vyg,SDSS:2004kqt} to baryon acoustic oscillations (BAO) and cosmic microwave background (CMB) measurements, has consistently confirmed that the universe is undergoing a phase of accelerated expansion. However, the nature of this acceleration, attributed to an exotic form of matter known as dark energy (DE), along with its fundamental origin, remains inadequately understood. Conversely, observational efforts to study potential late-time cosmological evolution have generated significant interest, as they may both reveal DE's fundamental nature and help resolve the Hubble tension ($H_0$ tension) \citep{Vagnozzi:2023nrq,Poulin:2024ken,Simon:2024jmu}. Consequently, the potential transitions in the dynamics of the universe related to the characteristics of DE have long been of significant interest for observational studies.
	
	A key component of DE is its equation of state (EoS) parameter, referred to as $w$, which is defined as the ratio of pressure $p$ to its energy density $\rho$. This parameter is essential for identifying the distinct characteristics of DE. One of the simplest methods to describe DE is by treating it as a perfect fluid, characterized by either a constant $(\Lambda)$ or a redshift-dependent EoS parameter. In instances where the EoS varies, a specific functional form of $w(a)$ is typically defined \citep{Linder:2002et,Caldwell:2005tm,Linder:2024rdj2,Park:2024vrw}. This function may be expressed as a Taylor series in terms of redshift, scale factor, or through a more generalized parameterization method. The associated parameters can subsequently be constrained with various observational datasets.
	At a cosmological level, distinguishing among different DE models can be achieved by examining the functional form of $w(a)$ along with the observational data of SNeIa \citep{Scolnic:2021amr}, CMB \citep{Planck:2018vyg} and BAO \citep{DESI:2024mwx,Chaussidon:2025npr}. To facilitate these comparisons, numerous parametrizations of $w(a)$ have been developed, with the majority expressed using two parameters.  

Numerous parameterizations of the DE EoS have been proposed in the literature, among which the Chevallier--Polarski--Linder (CPL) form is one of the most widely used. It is defined as $w_{\rm de} = w_0 + w_1(1 - a) = w_0 + \frac{w_1 z}{1 + z}$ and was originally introduced in \citep{Chevallier:2000qy, Linder:2002et}. The CPL form can be interpreted as a first-order Taylor expansion around the present-day scale factor ($a_0 = 1$). Despite its popularity, the CPL model often yields poor performance, particularly resulting in large uncertainties on the parameter $w_1$ \citep{Colgain:2021pmf}. Additional discussions on this issue can be found in \citep{Paliathanasis:2025cuc,Scherer:2025esj, RoyChoudhury:2025dhe,Arora:2025msq}. Importantly, the degree of this sensitivity is governed by the specific functional form that multiplies the parameter $w_1$ in the standard two-parameter EoS formulation $(w_0, w_1)$. To capture more complex behavior of DE, this model can be extended by including a second-order term $w_{\rm de} = w_0 + w_1(1 - a) + w_2(1 - a)^2 \cdots$. This generalization improves the model’s capacity to describe the time variation of the EoS, especially at low redshift \citep{SDSS:2004kqt}.
	\par  An alternative to the CPL form is the Pad\'{e} parameterization, which is mathematically robust and remains finite across all future redshifts \citep{Rezaei:2017yyj}. In all limiting cases, the EoS parameter asymptotically approaches constant values. These asymptotic values, particularly those corresponding to the early and present epochs of the Universe, may coincide with the cosmological constant value $w_\Lambda = -1$ or deviate from it, depending on how the observational data constrain the model parameters. By examining the evolution of the EoS from its early-time limit, given by $\frac{w_0 + w_1}{1 + w_2}$, to its present value $w_0$, one can assess whether the EoS significantly deviates from $w_\Lambda = -1$. A common feature of all parametric models is their limited sensitivity to dynamical DE at low redshift ($z\approx 0$), with this sensitivity gradually increasing at higher redshifts.

	As an illustration, consider the general form of the EoS parametrization expressed as $w_{\rm de}(z) = w_0 + w_1 X(z)$ \citep{Colgain:2021pmf}. It is demonstrated that the precision with which the parameter $w_1$, which governs the variation of the EoS, can be constrained from observational data is critically dependent on the behavior of the function $X(z)$ that modulates its evolution. A more rapid increase in $X(z)$ with redshift typically results in tighter constraints on $w_1$, whereas a slower growth leads to greater uncertainties. This inverse correlation between $w_1$ and the growth rate of $X(z)$ may pose challenges in interpreting dynamical DE parameterizations.
	
	This encourages us to explore an alternative parameterization known as generalized emergent DE (GEDE), which incorporates both the cold dark matter (CDM) model and the phenomenological emergent DE (PEDE). The GEDE framework incorporates two novel parameters: $w_{0}$, which denotes the specific model under consideration, and a transition redshift $z_t$. This latter parameter is defined by the condition $\Omega_{de}(z_t) = \Omega_m(z_t)$, thereby establishing a relationship between the matter density parameter and the transition redshift $z_{t}$, indicating that $z_{t}$ is not an independent variable. This GEDE model provides a broader cosmological perspective that connects the $\Lambda$CDM model with the PEDE model, and any deviation from $w_{0}= 0$ \footnote{It must not be confused with the parameter $w_0$ used in the GEDE model introduced in Eq.(\ref{eos3}) and the ones used in the CPL model and its generalizations introduced in Eqs.(\ref{eq2}) and (\ref{eq7}).}.  will enhance the plausibility of the new framework. 
	The model has been evaluated by various researchers through the Bayesian evidence analysis, which indicates a preference for the GEDE model over the $\Lambda$CDM across most observational datasets \cite{Yang:2021eud,Hernandez-Almada:2020uyr}. 
	
	We would like to underscore that the flexibility and generality we offer are particularly significant for our research on $w(z)$. These attributes not only expand the range of possibilities available for testing but also mitigate the potential for misleading results that can arise from some particular EoS. Henceforth, we consider another form which is characterized by a dimensionless parameter $\beta$. In the limits $\beta$ tending to $(-1, 0, +1)$, this EoS form fully generalizes three of the most common EoS parameterizations investigated in the literature and admits a much wider range of solutions (For more details on this, see \cite{Barboza:2009ks}). 
	
	Across all models considered in this work, we propose a novel parameterization referred to as ``BellDE", inspired by a Gaussian-type EoS. Models with more than two free parameters, particularly those that govern high-redshift behavior, often suffer from weak observational constraints. In such cases, conclusions suggesting phantom-like behavior may be premature or misleading. To address this, the BellDE model introduces a bell-shaped EoS with approximately two effective parameters. This form prevents phantom crossing at high redshifts while retaining sufficient flexibility in the low-redshift regime, where DE plays a dominant role in cosmic acceleration.
	
	In this study, we explore several parameterizations of $w(a)$ to deepen our understanding of DE dynamics, using observational data from DESI DR2 \citep{DESI:2025wyn,DESI:2025zpo}. We combine the DESI DR2 BAO data with Type Ia supernova measurements from Pantheon+ and DES5YR \citep{Brout:2022vxf,DES:2024jxu}, enabling a comprehensive analysis of DE characteristics across a wide redshift range.
	
	The manuscript is organized in the following way: Section \ref{sec2} introduces the theoretical framework concerning four distinct parameterizations of the EoS and their key physical attributes. Section \ref{sec3} outlines the methods and datasets used in our analysis. The findings are also detailed in the same section. A new bell-shaped EoS, which we propose, is explored in Section \ref{sec4}. Further, we conclude our studies with a summary in Section \ref{sec5}.
	
	\section{DE EoS  parametrizations}
	\label{sec2}
	
	In a universe that is spatially flat, homogeneous, and isotropic, the Friedmann constraint equation can be expressed as (ignoring the contribution from radiation $\Omega_{r0}$):
	\begin{equation}
		\frac{H^2(a)}{H_0^2} = \Omega_m^{(0)} a^{-3} + \left(1 - \Omega_m^{(0)}\right) \exp\left[3 \int_a^1 \frac{1 + w(a')}{a'} \, da' \right]\ ,
	\end{equation}
	where $H_0$ denotes the present-day Hubble parameter, and the current matter density parameter is defined as $\Omega_m^{(0)} = \frac{8\pi G \rho_m^{(0)}}{3H_0^2}$. Consequently, the present-day DE density parameter is given by $\Omega_{\text{DE}}^{(0)} = 1 - \Omega_m^{(0)}$, assuming spatial flatness. The scale factor evolves with redshift as $a = \frac{1}{1+z}$.
	
	In this framework, the underlying theory of gravity remains unmodified, enabling precise reconstruction of key cosmological parameters through various parameterizations of the DE EoS as outlined below:
	\begin{itemize}
		\item \textbf{Extended Chevallier-Polarski-Linder (CPL):} To begin with, we analyze a parameterized approach by expressing the time-dependent $w(a)$ for DE using a Taylor series expansion around $a_0 =1$ \citep{Dai:2018zwv}:
		\begin{equation}
        \label{eq2}
			\omega_{\rm de}(a)  = \omega_{0} + (1-a) \omega_{1} + (1-a)^2 \omega_2 + \cdots \ ,
		\end{equation}
		where the approximation has been performed up to the second order. In contrast to the first-order approximation, we extend the expansion to second order in the scale factor $a$. This approach has been taken to tackle the recent issue that emerged from the DESI DR2 data, where DE becomes a phantom at higher redshifts. 
		
		\item \textbf{PADE approximation:} Unlike the Taylor series approximation of the EoS  for DE, we now include a Pad\'{e} approximation of $\omega(a)$, which can be generally expressed as $(m/n)$ \citep{Wei:2013jya,Rezaei:2023xkj,Hu:2024qnx}
		\begin{equation}
			\omega_{\rm de}(a) \approx \frac{p_0 + p_1 (a - 1) + \cdots + p_m (a - 1)^m}{1 + q_1 (a - 1) + \cdots + q_n (a - 1)^n} \ .
		\end{equation}
		Truncating the series up to first order (1/1), we get: 
		\begin{equation}
			\omega_{\rm de}(a) \approx \frac{\omega_{0} + \omega_{1}(1-a)}{1 + \omega_{2} (1-a)} \ .
		\end{equation}
		It is important to note that various DE equations of state proposed in the literature can be recovered as special cases of this general parametrization.
		
		\item \textbf{GEDE:} The EoS for the GEDE model is \citep{Hernandez-Almada:2020uyr}: 
		\begin{multline}
		\omega_{\rm de}(z) = -1 - \frac{w_{0}}{3 \ln 10} 
		 \times \left[1+ \tanh \left(w_{0} \times \log_{10}\left(\frac{1+z}{1+z_t}\right)\right)\right],
		\label{eos3}
		\end{multline}
		where $z_t$ denotes the transition redshift, determined by the condition $\Omega_{\text{DE}}(z_t) = \Omega_{\rm m} (1+z_t)^3$, and therefore it is not treated as a free parameter. In terms of the scale factor, the above expression can be rewritten as: 
		\begin{equation}
		\omega_{\rm de}(a) = - 1- \frac{w_{0}}{3 \ln 10} 
		 \times \left[1+ \tanh \left(w_{0} \times \log_{10}\left(\frac{a_t}{a}\right)\right)\right].
		\end{equation}
		It is observed that when $w_0 = 0$, the standard $\Lambda$CDM model is regained, and in the case where $w_0 = 1$ and $z_t = 0$, the  PEDE model is obtained \citep{Li:2019yem,Liu:2024wne,Wang:2024dcx,Yang:2021eud}.
		
		\item \textbf{Generalized DE (GDE):} Here, we investigate another interesting parameterization characterized by a dimensionless parameter $\beta$ \citep{Barboza:2009ks}. In the limits of $\beta$, this formulation provides a comprehensive generalization of three of the most prevalent EoS parameterizations examined in the literature and accommodates a considerably broader range of solutions.
		
		The EoS is given by: 
		\begin{equation}
        \label{eq7}
			\omega_{\rm de}(a) = \omega_{0} - \omega_{1} \frac{a^{\beta} - 1}{\beta} \ .
		\end{equation}
		The choice of $\beta$ leads to diverse representations of the EoS. For instance,
		\begin{equation}
			\omega_{\rm de}(a) = 
			\begin{cases}
				\omega_{0} + \omega_{1} \left(\frac{1}{a}-1\right), & \beta =-1,\\
				\omega_{0} + \omega_{1} \ln a^{-1}, & \beta = 0,\\
				\omega_{0} + \omega_{1} (1-a), & \beta =1\ .
			\end{cases}
		\end{equation}
		In this study, we vary the parameter $\beta$ and impose constraints based on the considered data.
		
	\end{itemize}
	
	
	\section{Methodology and Data} \label{sec3}
	
A thorough parameter estimation of cosmological parameters is performed using a Markov Chain Monte Carlo (MCMC) approach. To gain deeper insight into the nature of DE, the following observational datasets are considered:
	\begin{itemize}
		\item Cosmic Chronometers \textbf{(CC)}: Measurements of the Hubble parameter are derived from the CC approach \citep{Yu:2017iju}. The function $H(z)$ can be determined by computing the derivative of the cosmic time with respect to the redshift at $z \neq 0$. To find the rate of change $\frac{\Delta z}{\Delta t}$, it is necessary to assess the age difference between two galaxies at different redshifts.
		
		\item   \textbf{BAO}: We utilize BAO measurements from DESI's second data release, which includes observations of galaxies and quasars, as well as Lyman-$\alpha$ tracers. These measurements cover both isotropic and anisotropic BAO constraints over $0.295 \leq z \leq 2.330$, divided into nine redshift bins. The BAO constraints are expressed in terms of the transverse comoving distance $D_M/r_d$, the Hubble horizon
		$D_H/r_d$, and the angle-averaged distance $D_V/r_d$, all normalized to the comoving sound horizon at the drag epoch, $r_d$. This dataset is referred to as DESI-DR2 \citep{DESI:2025zgx}.
		
		
		\item Supernovae Type Ia (SNeIa):  \textbf{Pantheon+ (PP)-} The distance modulus
		measurements from SNeIa in the Pantheon+ sample include 1701 light curves from 1550 distinct SNeIa events, spanning a redshift range of $0.01$ to $2.26$ \citep{Brout:2022vxf,Scolnic:2021amr}. We have ignored the SH0ES calibration for this analysis and using the observational column corresponding to the apparent magnitude $m$ \footnote{The distance modulus is defined as $\mu \equiv m-M_b = 5\log_{10}(D_L/\text{Mpc}) + 25$, where $M_b$ denotes the absolute brightness of the Type Ia Supernova, and \(D_L\) is the luminosity distance.}. \\
		\textbf{DESY5-} As part of their Year 5 data release, the DE Survey (DES) recently published results from a new, homogeneously selected sample of 1635 photometrically classified SN Ia with redshifts spanning $0.1 < z < 1.3$ \citep{DES:2024jxu}. This sample is complemented by 194 low-redshift SN Ia (shared with the PantheonPlus sample) in the range $0.025 < z < 0.1$. We refer to this samples as `DES'. The likelihood is estimated by marginalizing $M_B$ using the python code given in \href{https://github.com/des-science/DES-SN5YR/blob/main/5_COSMOLOGY/SN_only_cosmosis_likelihood.py}{DES-SN5YR module}. The process of analytically marginalizing $M_B$ is given in the ref. \citep{Goliath:2001af}.
		
		\item   \textbf{CMB}: We use the compressed CMB data from distance priors, which contain essential information from the power spectrum. The parameters included in this study are:
		\begin{itemize}
			\item The acoustic scale ($l_A$), characterizing the CMB temperature power spectrum in the transverse direction.
			\item The shift parameter ($R$), governing the CMB temperature spectrum along the line of sight.
		\end{itemize}
		The data for these parameters, along with their correlation matrix, are obtained from \citep{Chen:2018dbv}. We refer to this dataset as ``PLANCK".
		
	\end{itemize}
	
	\begin{table}
    \centering
		\begin{tabular}{|lrr|}
			\hline
			\hline
			Parameters & Prior Range &\\
			\hline
			$\Omega_{m}$ & [0, 1.0] & \\
			$H_0$  & [30,100] & \\		
			$w_{0}$ & [-2,2] &\\
			$w_{1}$ & $[-6.5, 2.0]$ & \\
			$w_{2}$ & $[-3.0,6.5]$ & \\
			$\beta$ & $[-2.5,2.5]$ & \\
			$M_b$ & $[-20, -18]$ & \\
			\hline
			\hline
		\end{tabular}
		\caption{The uniform prior ranges for the model parameters, including (CPL, PADE, GEDE, GDE). }
		\label{tab:priors}
	\end{table}

	\begin{table*}
		\begin{tabular} { l  c c c c c c c }
			\noalign{\vskip 3pt}\hline\hline \noalign{\vskip 5pt}
			\multicolumn{1}{c}{} &  \multicolumn{1}{c}{\bf BASE+PP} &  \multicolumn{1}{c}{\bf BASE+DES} & \multicolumn{1}{c}{\bf BASE+PP} &  \multicolumn{1}{c}{\bf BASE+DES} &  &{\bf BASE+PP} & {\bf BASE+DES}\\
			\noalign{\vskip 3pt}\cline{1-8}\noalign{\vskip 3pt}
			
			Parameters &  68\% limits &  68\% limits &  68\% limits &  68\% limits & Parameters &  68\% limits &  68\% limits\\
			\hline
			\multicolumn{3}{c}{\textbf{CPL}} & \multicolumn{2}{c}{\textbf{PADE}} &  \multicolumn{3}{c}{\textbf{GDE}} \\ 
			\hline   \\[-1.5ex] \rule{0pt}{2ex}  
			{\boldmath$\Omega_{m} $} & $0.3195^{+0.0023}_{-0.0021}$  &  $0.3195^{+0.0024}_{-0.0021}$ & $ 0.3178 \pm 0.0023$ &  $ 0.3180\pm 0.0023$ & {\boldmath$\Omega_{m} $} &  $0.3145 \pm 0.0023$ & $0.3144\pm 0.0023$  \\    [0.5ex]
			
			{\boldmath$H_0 $} & $67.7 \pm 1.6$ & $68.63 \pm 0.30$ & $67.9 \pm 1.6$ & $68.56 \pm 0.30$ & {\boldmath$H_0 $} & $68.2 \pm 1.6$ &  $68.79 \pm 0.29$ \\ [0.5ex]
			
			{\boldmath$w_{0}    $} & $-0.798^{+0.075}_{-0.084}$ & $-0.812^{+0.074}_{-0.088}$ & $-0.714 \pm 0.072$ & $-0.720^{+0.060}_{-0.073}$ & {\boldmath$w_{0}    $} & $-0.809^{+0.044}_{-0.053}$ & $-0.793^{+0.046}_{-0.062}$ \\ [0.5ex]
			
			{\boldmath$w_{1}  $} & $-0.39^{+0.88}_{-0.63}$ & $-0.25^{+0.86}_{-0.59}$ & $-3.16^{+0.89}_{-2.3}$ & $-2.6 \pm 1.6$ & {\boldmath$w_{1}$} & $ -0.63^{+0.35}_{-0.22}$ & $-0.71^{+0.47}_{-0.21}$\\ [0.5ex]
			
			{\boldmath$w_{2} $} & $-0.9^{+1.0}_{-1.6}$ &  $-1.09^{+0.87}_{-1.5}$ &   $ 2.1^{+1.9}_{-1.1}$ &    $1.5^{+1.4}_{-1.6}$ & {\boldmath$\beta    $} & $0.24^{+1.4}_{-0.88}$ &  $0.3^{+1.2}_{-1.7}$ \\ [0.5ex]
			{\boldmath$M_b$} & $-19.40 \pm 0.053$ &  $-$ &   $-19.393 \pm 0.052$ &    $-$ &  {\boldmath$M_b$} &$-19.40 \pm 0.052$&   --                 \\ [0.5ex]
			
			\hline  \vspace{0.05in}
			{\boldmath$\chi_{red}^{2}$} & $1.019$ &  $0.899$ &   $ 1.019$ &  $0.899$ & & $1.019$ &  $0.899$ \\ [0.5ex]
			{\boldmath$AIC$} & $1781.71$ &  $1686.66$ &   $ 1781.74$ &  $1686.71$ & & $1782.83$ &  $1688.89$ \\ [0.5ex]
			
			\hline

			\multicolumn{3}{c}{\boldmath \bfseries BellDE} & \multicolumn{2}{c}{\boldmath \bfseries GEDE} & \multicolumn{3}{c}{} \\
			\hline
			
			{\boldmath$\Omega_{m}  $} & $ 0.3139\pm 0.0022$  & $0.3148\pm 0.0021$ & $ 0.3087 \pm 0.0018$ & $0.3082\pm 0.0018$\\[0.5ex]
			
			{\boldmath$H_0 $} & $68.3\pm 1.7 $ & $68.94\pm 0.28$ & $ 68.7 \pm 1.7$ & $69.50 \pm 0.23$ & & & \\[0.5ex]
			
			{\boldmath$w_{0}     $} & $-1.48^{+0.18}_{-0.21}$ & $-1.73\pm 0.26$& $-0.264 \pm 0.070$ & $-0.289 \pm 0.069$ & & \\[0.5ex]
			
			{\boldmath$w_{1}  $} &    $-0.826^{+0.050}_{-0.093}$    &  $ -0.841^{+0.037}_{-0.069}$ &&&& &\\[0.5ex]
			
			{\boldmath$z_{t}           $} & $1.65^{+0.29}_{-0.34}$ &  $1.81\pm 0.30$  &&&&&                         \\[0.5ex]
			{\boldmath$\Delta         $} &    $0.967\pm 0.097$    &   $0.960\pm 0.099$   &&&&&                       \\[0.5ex]
			{\boldmath$M_b$} & $-19.390\pm 0.053$ &  $-$ &            $-19.391 \pm 0.053$ & --& & &        \\ [0.5ex]
			\hline
			\vspace{0.05in}
			{\boldmath$\chi_{red}^{2}$} & $1.020$ &  $0.900$ &   $ 1.027$ &  $0.909$ &  &  \\ [0.5ex]
			{\boldmath$AIC$} & $1783.30$ &  $1689.12$ &   $ 1792.73$ &  $1703.36$ & &  \\ [0.5ex]
			
			\hline
			
			\multicolumn{3}{c}{\bf\boldmath Flat \(\Lambda\)CDM} & & & &  & \\
			\hline
			
			{\boldmath $\chi_{red}^2$} & $1.033$ & $0.917$ & &&&&\\[0.5ex]
			{\boldmath $AIC$} & $1803.76$ & $1717.314$ & &&&&\\
			\hline
			\hline
		\end{tabular}
		\caption{Cosmological constraints on the models: CPL, PADE, GDE, GEDE, BellDE based on the BASE+PP and BASE+DES, where BASE refers to the data combination CC+DESBAO+PLANCK. }
		\label{tab:best_fit}
	\end{table*}
	
	We conduct MCMC analyses utilizing the publicly accessible Python package emcee3 to refine the parameter space of the underlying models. In our parameter inference, we select a wide uniform prior range tabulated in Tab.  \ref{tab:priors}. The obtained samples were analyzed with the well-known GetDist package \citep{Lewis:2019xzd}, and the best-fit parameters of the models are listed in Tab. \ref{tab:best_fit}. 
	
	In particular, we examine the constraints on model parameters using combinations of the DESI DR2 dataset with other observational probes. Specifically, we consider two data combinations: (i) CC + DESI BAO + Planck + Pantheon+, and (ii) CC + DESI BAO + Planck + DESI Y5. The combination CC + DESI BAO + Planck serves as a consistent Baseline across both cases.
	
\begin{figure}
		\centering
		\includegraphics[scale=0.54]{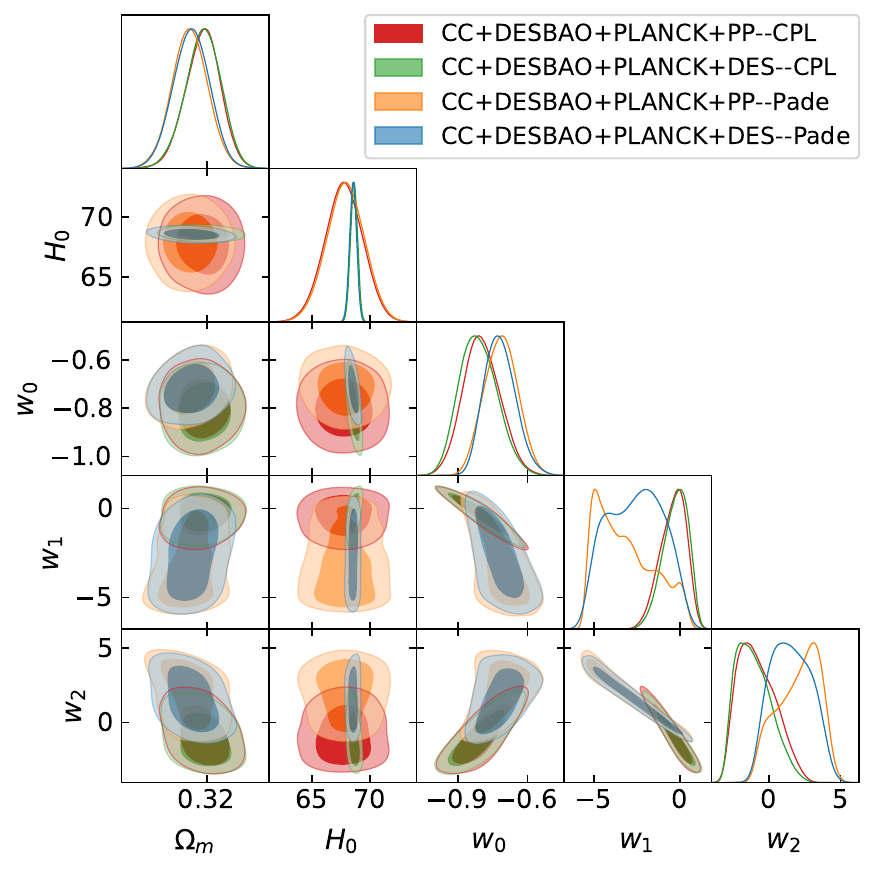}
		\includegraphics[scale=0.4]{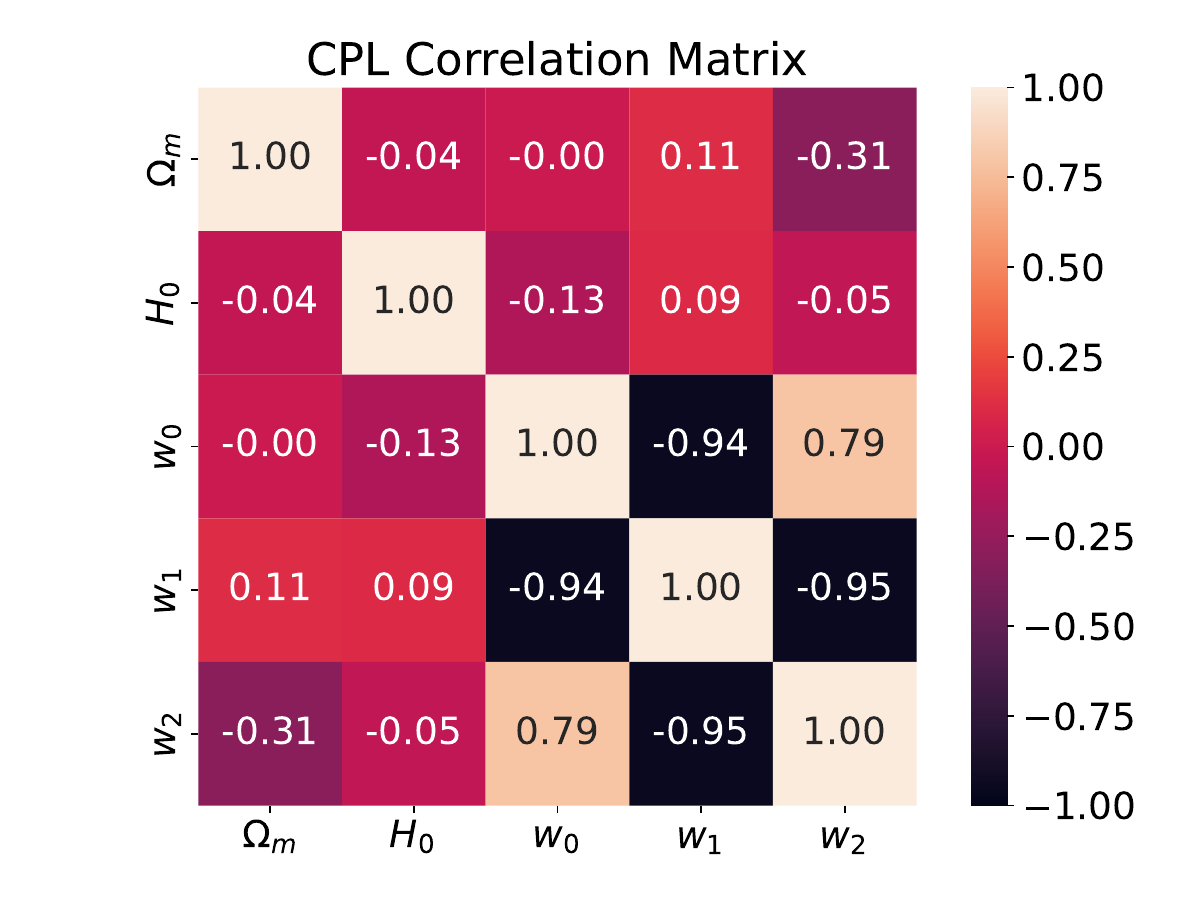}
		\includegraphics[scale=0.4]{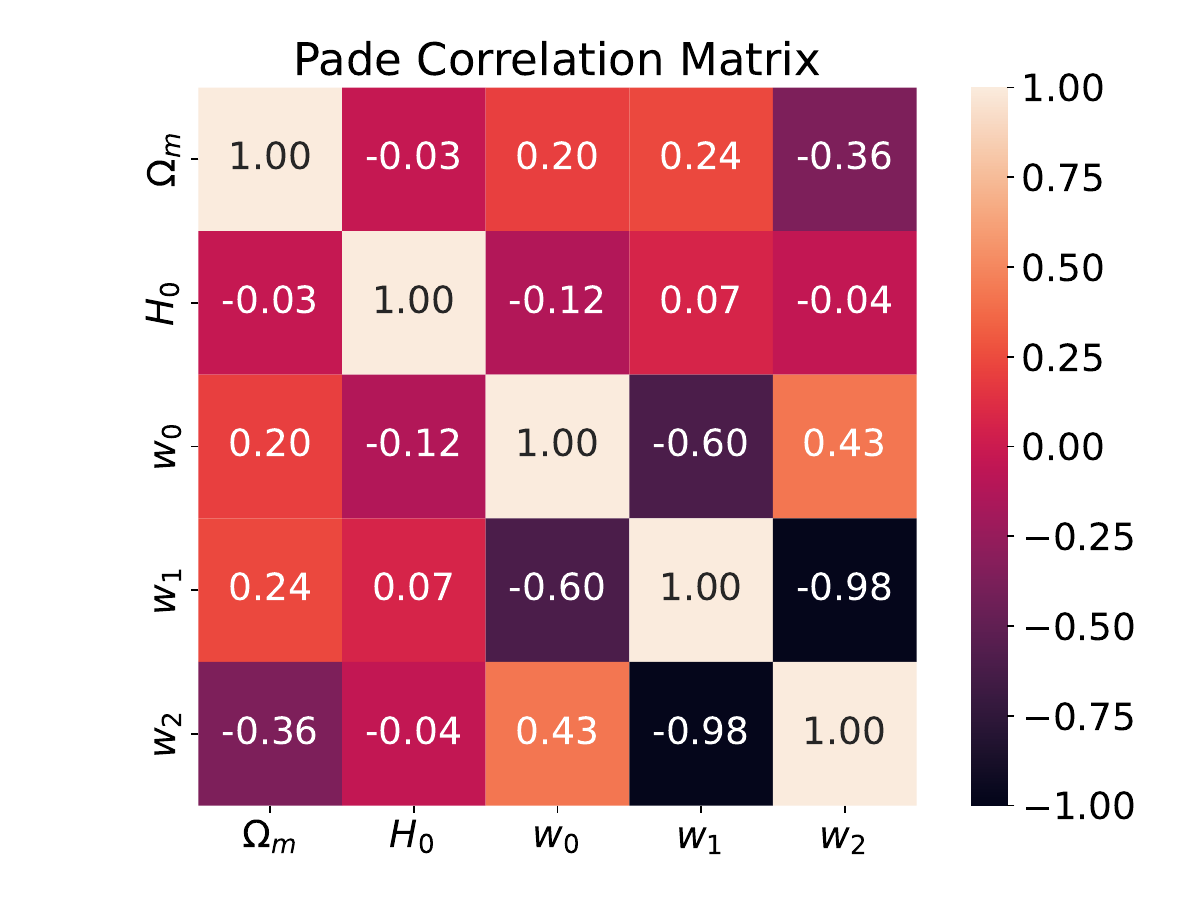}
		\caption{The 1D and 2D marginalized posterior distributions of the model parameters for the CPL and Padé parameterizations, with the nuisance parameter \(M_B\) marginalized over. The correlation matrix is constructed by combining both datasets and marginalizing over \(M_B\).}
		\label{fig:cpl}
	\end{figure}

\begin{figure}
		\centering
		\begin{minipage}{0.9\linewidth}
			\centering
			\includegraphics[width=\linewidth]{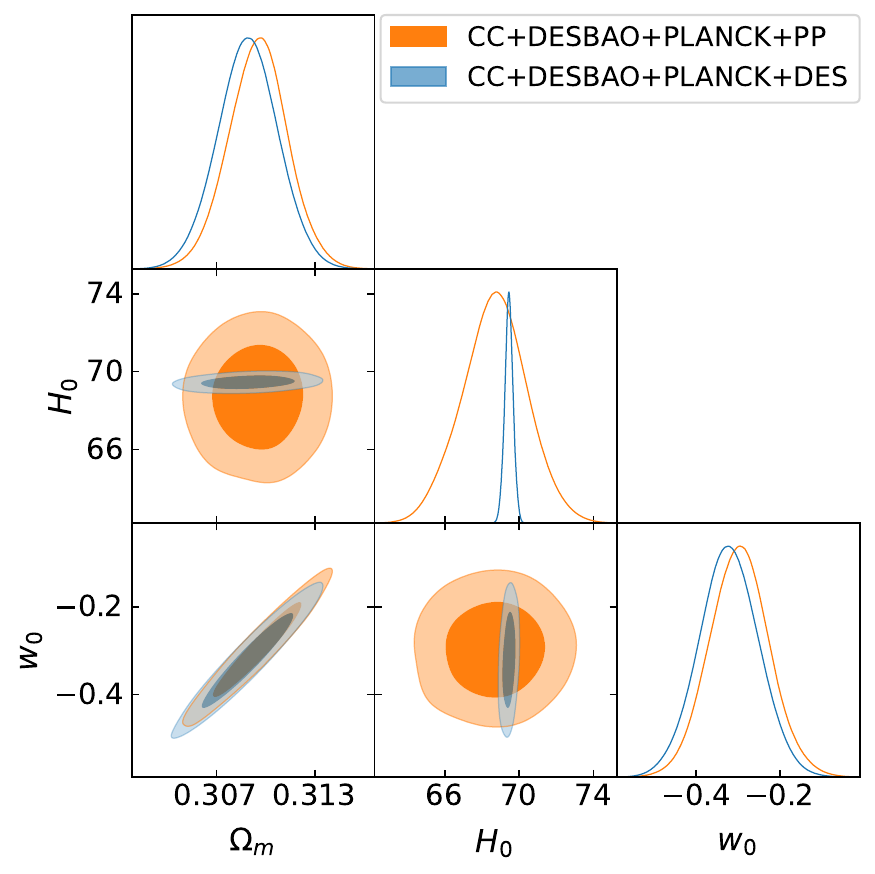}
			
		\end{minipage}
		\vspace{0.3cm}
		
		\begin{minipage}{0.9\linewidth}
			\centering    
			\includegraphics[width=\linewidth]{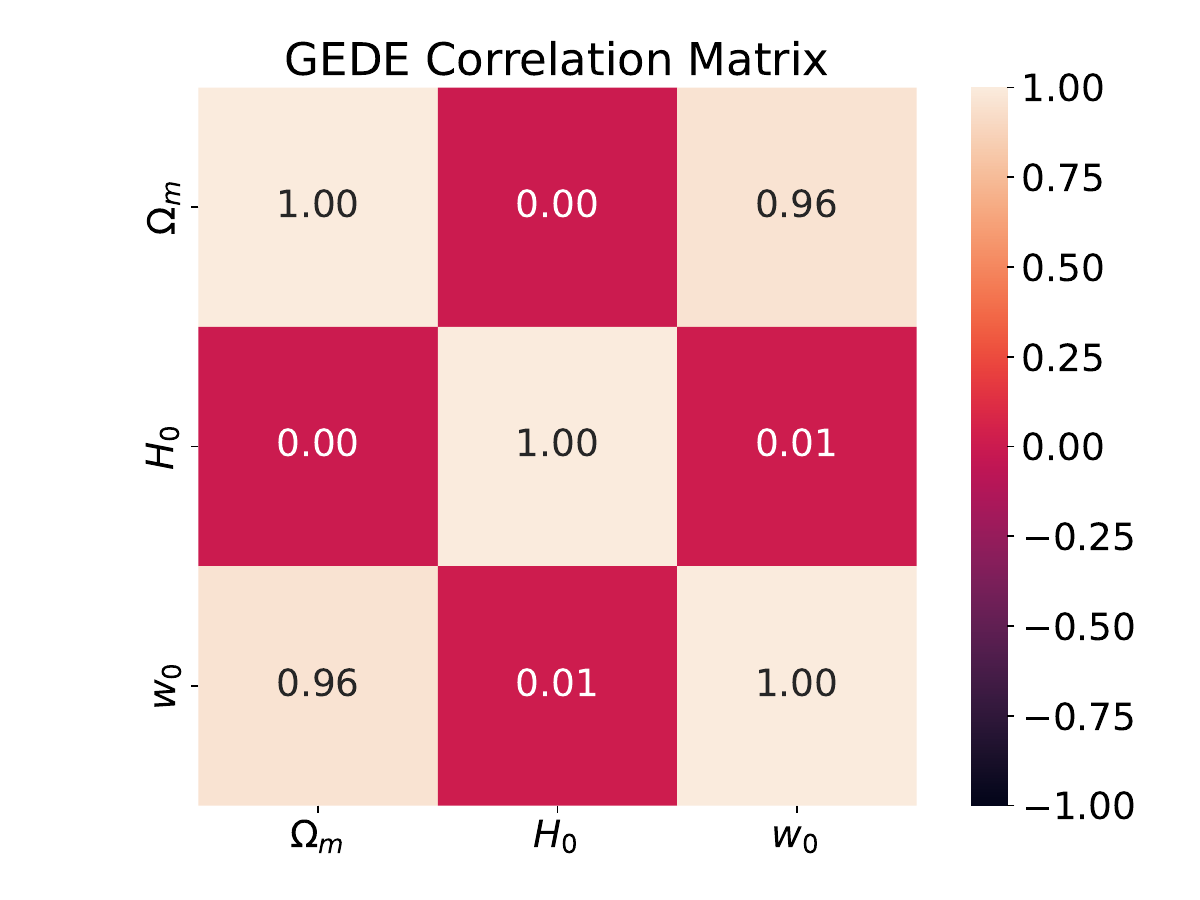}
		\end{minipage}
		\caption{The 1D and 2D marginalized posterior distributions of the model parameters for the Generalized Emerging DE (GEDE) parameterization, with the nuisance parameter \(M_B\) marginalized over. The correlation matrix is constructed by combining both datasets and marginalizing over \(M_B\).}
		\label{fig:gede}
	\end{figure}

\begin{figure}
		\centering
		\begin{minipage}{0.9\linewidth}
			\centering
			\includegraphics[width=\linewidth]{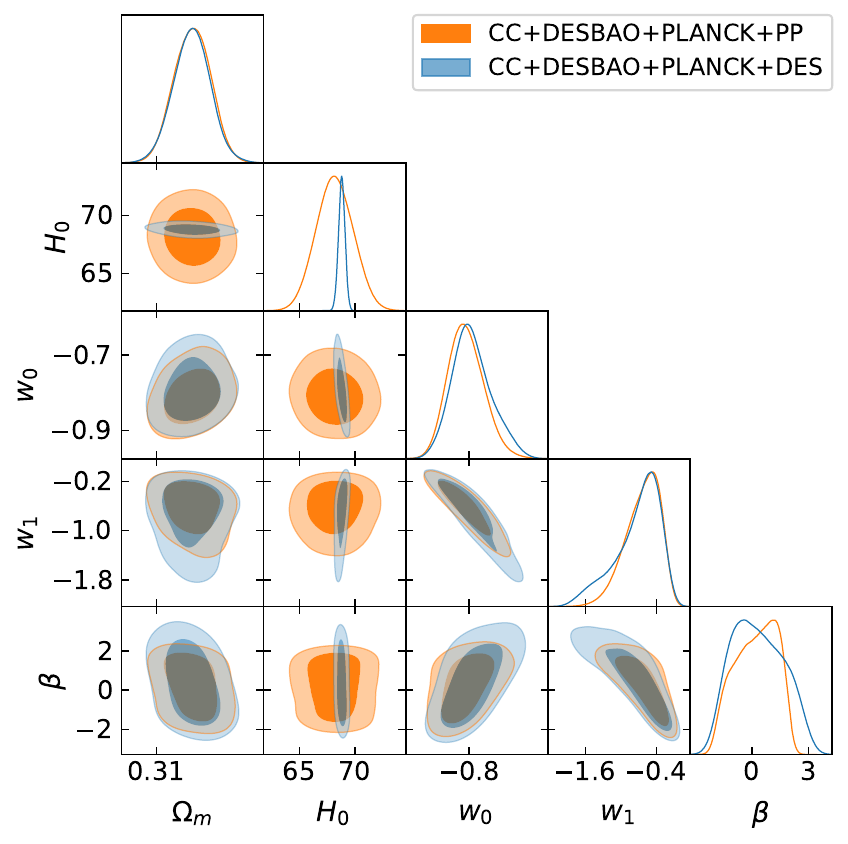}
		\end{minipage}
		\vspace{0.3cm}
		\begin{minipage}{0.9\linewidth}
			\centering
			\includegraphics[width=\linewidth]{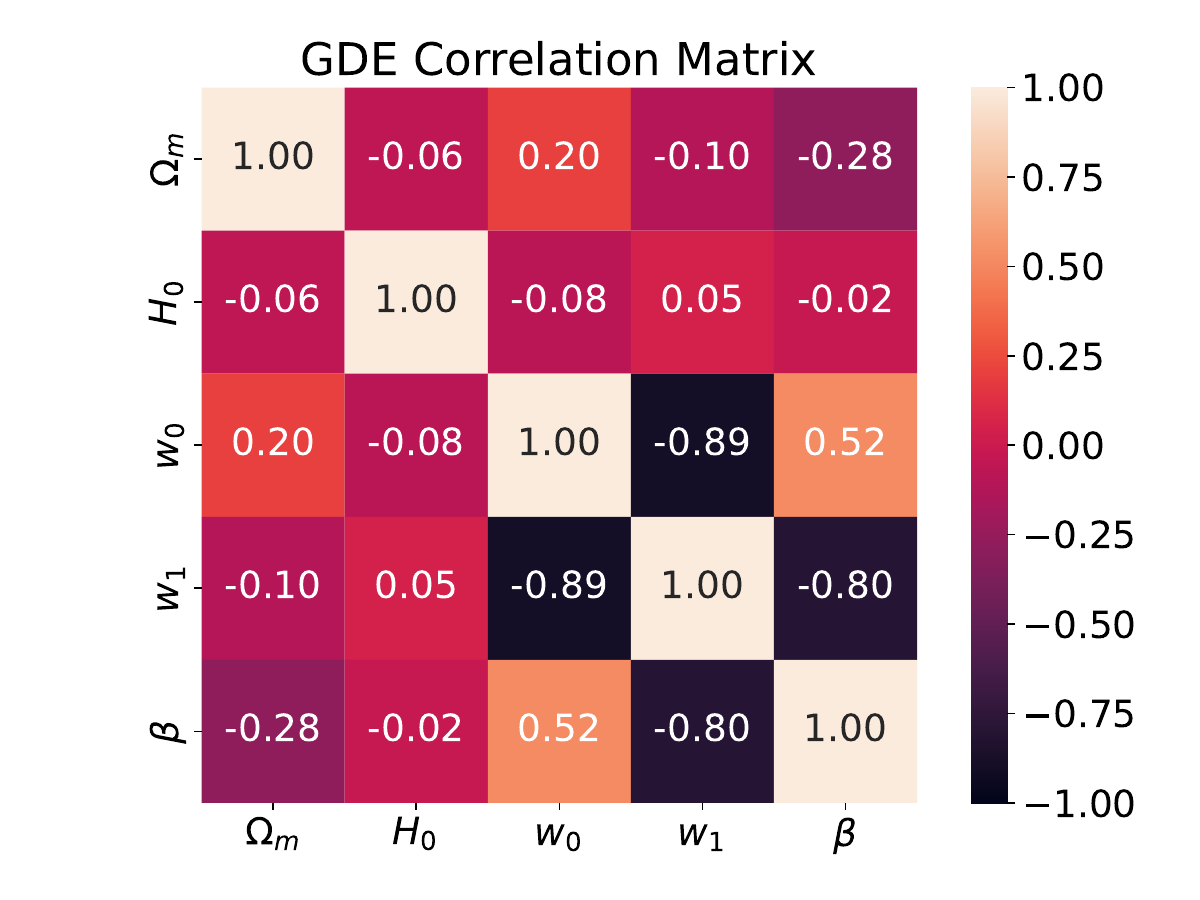}
		\end{minipage}
		\caption{The 1D and 2D marginalized posterior distributions of the model parameters for the Generalized DE (GDE) parameterizations, with the nuisance parameter \(M_B\) marginalized over. The correlation matrix is constructed by combining both datasets and marginalizing over \(M_B\).}
		\label{fig:gde}
	\end{figure}

	{Given the heterogeneity of the current SN datasets, some methodological care is warranted when combining them with other probes. The PP sample is based on apparent magnitude measurements, where the nuisance parameter $M_B$ is degenerate with the Hubble constant $H_0$. Therefore, in our analysis, external datasets (e.g., CMB and BAO) are crucial to breaking this degeneracy and obtaining meaningful cosmological constraints. In contrast, the DESY5 SN sample provides measurements of the distance modulus, where $M_B$ has already been calibrated independently, removing the direct degeneracy with $H_0$.\\
	Although both datasets are based on Type Ia supernovae, they differ in methodology, calibration, and treatment of nuisance parameters. As emphasized in recent studies \cite{Colgain:2024mtg, Dhawan:2024gqy}, combining supernova datasets without accounting for these differences can lead to tensions or biases in the inferred cosmological parameters. Moreover, the DES dataset includes many supernovae that are also present in the Pantheon+ compilation; thus, combining them without proper filtering would result in double counting and consequently introduce systematic inconsistencies. For these reasons, we refrain from combining PP and DESY5 together and instead analyze them separately in conjunction with the Baseline datasets. This approach ensures internal consistency and avoids potential systematic contamination from dataset mixing.}
	 Figures \ref{fig:cpl}, \ref{fig:gede}, and \ref{fig:gde} show a comparison of results for these two distinct data combinations. We also include the corresponding parameter correlation matrices, derived from the MCMC chains and visualized using the \texttt{Seaborn} Python package \citep{Waskom2021}.
	
\begin{figure*}
		\centering
		\includegraphics[scale=0.34]{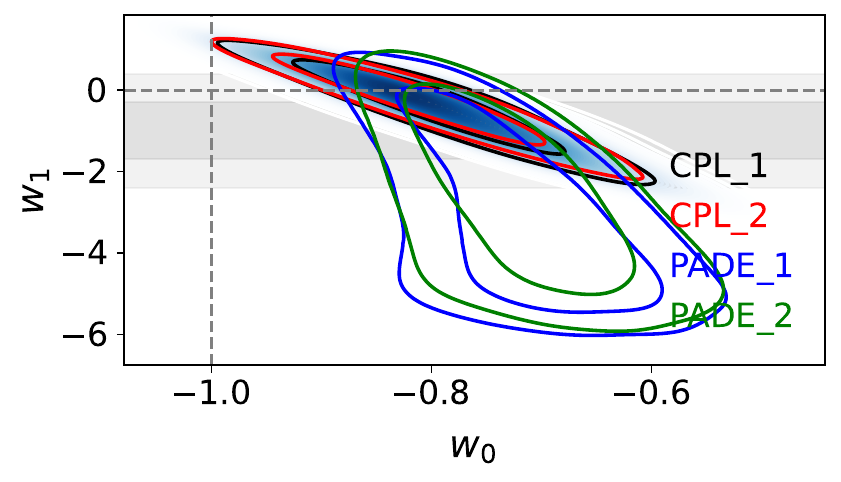}
		\includegraphics[scale=0.34]{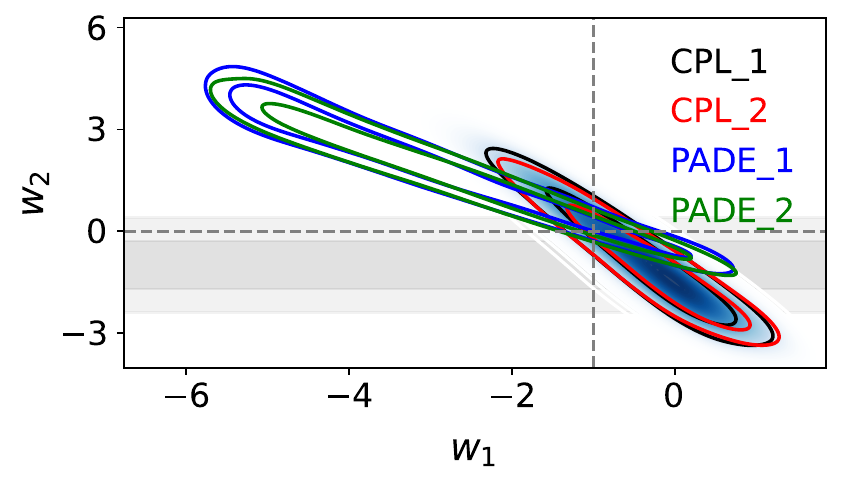}	
		\includegraphics[scale=0.34]{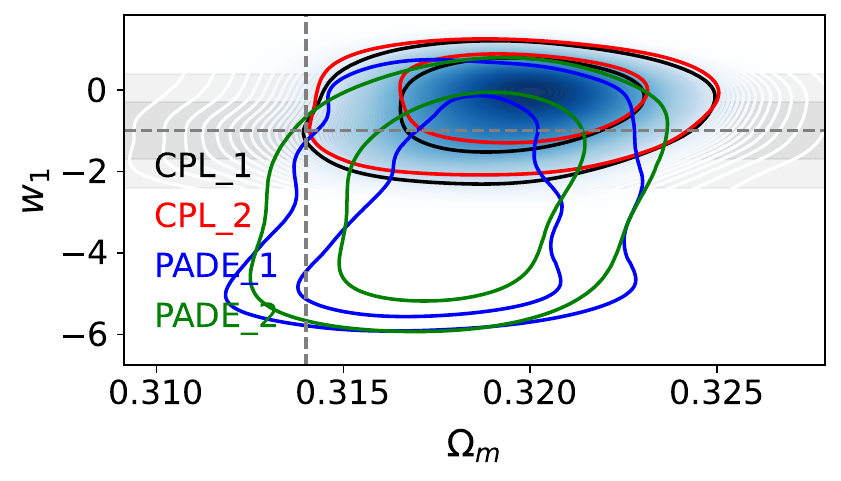}
		\caption{The 68\% and 95\% confidence contours in the \( w_0-w_1 \), \( w_1-w_2 \), and \( \Omega_{\rm M}-w_1 \) planes are shown using DESI DR2 BAO data in combination with Planck, CC, DES, and PP data. Here, \texttt{CPL\_1} and \texttt{CPL\_2} refer to the CPL parameterization applied to the BASE+PP and BASE+DES combinations, respectively; the same notation applies to the PADE model.}
		\label{fig:2d_posterior}
	\end{figure*}

\begin{figure}
		\centering
		\begin{minipage}{0.9\linewidth}
			\centering
			\includegraphics[width=\linewidth]{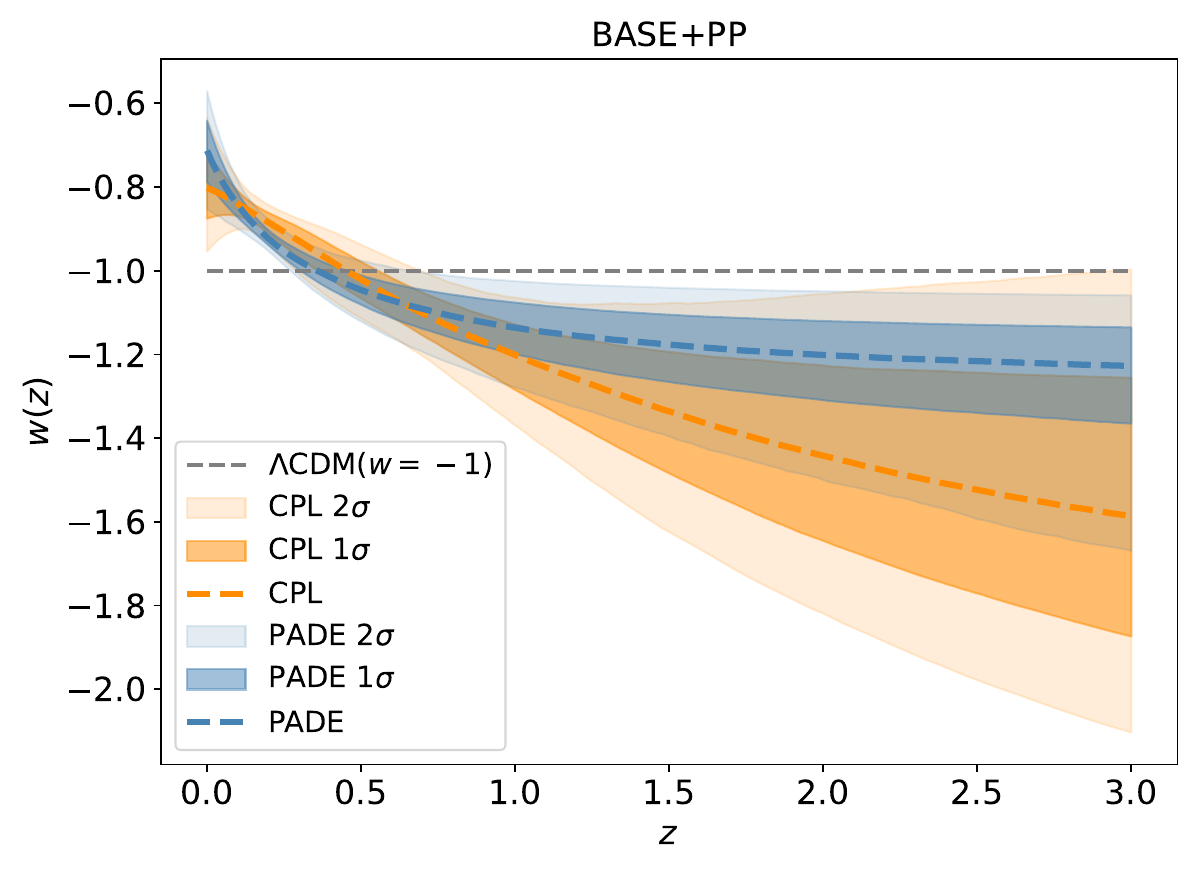}
		\end{minipage}
		
		\vspace{0.3cm}
		
		\begin{minipage}{0.9\linewidth}
			\centering
			\includegraphics[width=\linewidth]{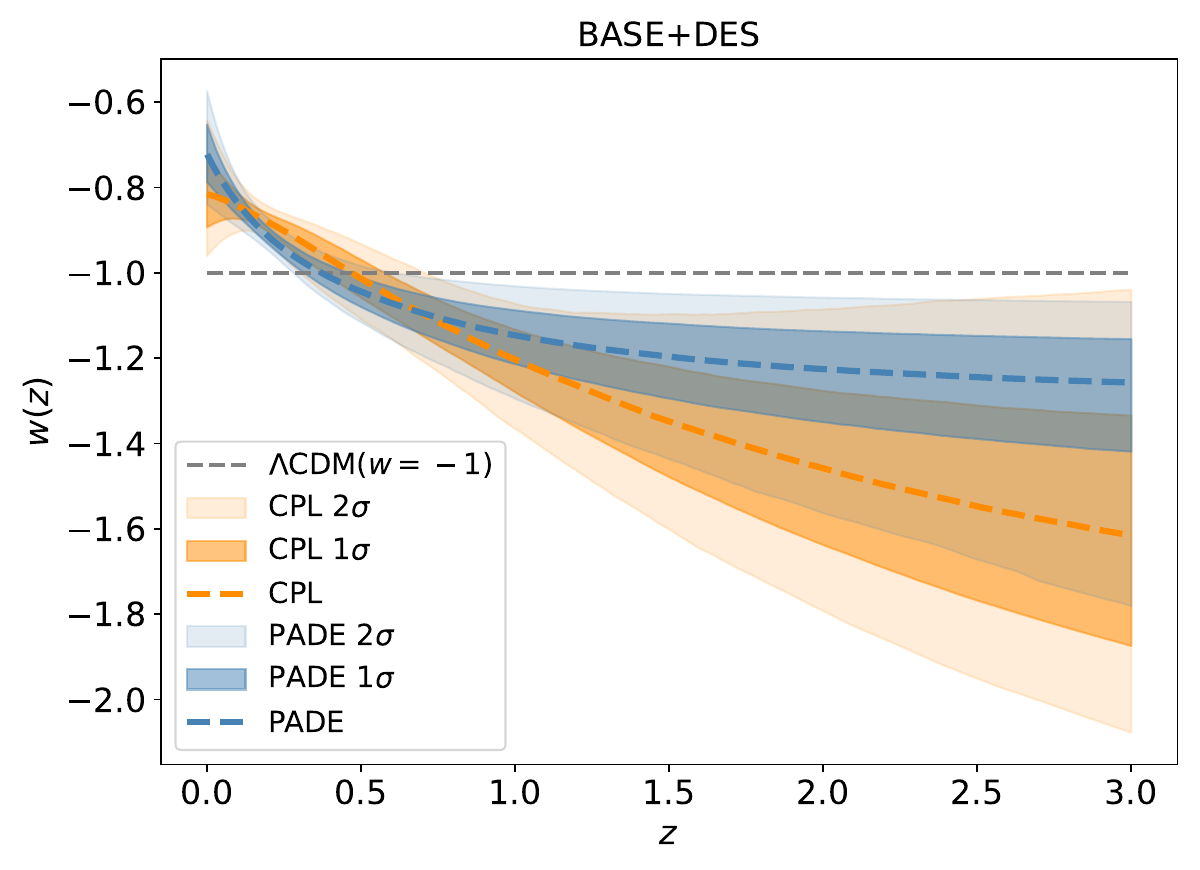}
		\end{minipage}
		
		\caption{Redshift evolution of the EoS $w(z)$. The upper panel illustrates the CPL and PADE models for the BASE+PP dataset, while the lower panel shows the results for BASE+DESY5. Shaded bands indicate the 68\%~ ($1\sigma$) and 95\%~($2\sigma$) confidence intervals derived from observational constraints. The dashed orange and blue curves represent the means of each model, and the horizontal grey dashed line corresponds to the standard $\Lambda$CDM curve.}
		\label{fig:w_cpl_pade}
	\end{figure}

	The best-fit values for the extended CPL model parameters are presented in Tab.~\ref{tab:best_fit}, and the corresponding MCMC chains with their correlation matrix are shown in Fig.~\ref{fig:cpl}. The combined dataset exhibits a preference for the region where $w_0 > -1$ and $w_1 < 0$, as illustrated in Fig.~\ref{fig:2d_posterior}. This behavior suggests a deviation from a cosmological constant, indicating that the EoS was phantom-like (\( w(z) < -1 \)) in the early Universe and has evolved to \( w(z) > -1 \) at present, as depicted in Fig.~\ref{fig:w_cpl_pade}.
	
It is important to note that the individual datasets may not, on their own, exhibit a strong preference for the phantom-crossing region of the parameter space. Previous Planck analyses within the $w_0w_a$CDM framework typically favor $w_0 > -1$ but do not place strong constraints on $w_a < 0$ unless combined with external datasets~\cite{Planck:2018vyg}. Similarly, combinations such as CC with PP do not show a significant indication of phantom behavior and impose only weak constraints on $w_a$. Despite these known differences, a consistent trend emerges when combining them, namely, BASE+PP and BASE+DES\footnote{By ``BASE", we refer to CC+DESIBAO+PLANCK}, with a preference for the region $w_0 > -1$, along with $w_1 < 0$ and $w_2 < 0$. In our analysis, a stronger preference for this region is observed when the DESI BAO data are included in combination with other datasets. This highlights the critical role of dataset combinations and the influence of known tensions between them, particularly around redshift \(z\sim0.5 \), as also discussed in~\cite{Colgain:2024xqj,DESI:2025fii}.
	
In this context, we perform a statistical comparison between our models and flat \(\Lambda\)CDM using the Akaike Information Criterion (AIC), defined as,
		\begin{equation}
			\mathrm{AIC} = 2k - 2\ln \mathcal{L}_{\mathrm{max}} \ ,
		\end{equation} 
		where \(k\) is the number of model parameters and \(\mathcal{L}_{\mathrm{max}}\) is the maximum likelihood. The AIC values for each model are summarized in Tab.~\ref{tab:best_fit}, alongside the reduced chi-squared statistic, defined as $\chi^2_{\rm red} \equiv \chi^2_{\rm min} / \nu$, where $\nu$ is the number of observational data points minus the number of fitted parameters. This statistic serves as a standard measure of goodness-of-fit: values of $\chi^2_{\rm red} \ll 1$ typically suggest overfitting, while $\chi^2_{\rm red} \gg 1$ indicates a poor fit. A value close to unity implies a good fit to the data. On the other hand, the Akaike Information Criterion (AIC) penalizes model complexity and rewards better fit, making lower AIC values indicative of a more favorable trade-off between goodness-of-fit and model parsimony.\\
		We fit the standard flat $\Lambda$CDM model using the same dataset combinations to serve as a benchmark for comparison. All models under consideration, including $\Lambda$CDM, yield reduced chi-squared values close to unity, confirming that each provides an acceptable fit to the data. However, the AIC values vary across models, revealing meaningful distinctions in statistical performance. Notably, the CPL parametrization yields lower AIC values compared to $\Lambda$CDM across all dataset combinations, suggesting a statistical preference for mild dynamical DE behavior within the CPL framework.\\
		In terms of parameter constraints, the combination including DESY5 SN data yields notably tighter bounds on the Hubble constant \(H_0\) compared to the PP combination. This indicates that the DESY5 data has a greater ability to break degeneracies, particularly those involving nuisance parameters like $M_B$. Nevertheless, the constraints on other cosmological parameters, such as \(\Omega_{\rm m}\) and \(w_0\), remain comparable across both combinations, with minor shifts in their central values. \\
		As we increase the number of free parameters in the EoS, such as by introducing $w_1$ and $w_2$, both dataset combinations lose constraining power, as reflected in the broader confidence contours shown in Fig.~\ref{fig:2d_posterior}. This underscores the difficulty of constraining higher-order terms in the DE parameter space with current data. Overall, these results highlight the enhanced sensitivity of the BASE+DES combination in constraining key parameters like \(H_0\), \(\Omega_{\rm m}\), and \(w_0\), while also emphasizing the inherent limitations in probing extended DE models.

	A similar behavior is observed in the Pad\'e parameterization, which exhibits significant phantom-like behavior at higher redshifts, while favoring \( w_0 > -1 \) at the present epoch. The triangular plots, along with the corresponding correlation matrix, are shown in Fig.~\ref{fig:cpl}. The 2D marginalized posterior distribution between the EoS parameters \( w_0 \) and \( w_1 \) is displayed in Fig.~\ref{fig:2d_posterior}, where we observe that even at the \(2\sigma\) level, a significant deviation of \( w_0 \) from \(-1\) is evident. Furthermore, the allowed range of \( w_2 \) shifts toward positive values, indicating an evolving EoS at late times.
	{Comparing the confidence regions of the CPL and PADE parametrizations, we observe that the CPL model exhibits broader $\sigma$ levels, particularly at higher redshifts, as shown in Fig. \ref{fig:w_cpl_pade}. This indicates weaker constraints on its parameters, likely due to its limited functional flexibility. In contrast, the PADE parametrization yields tighter bounds, suggesting that it offers a more effective and stable representation of the DE dynamics within the redshift range probed by the data.}
	
	Additionally, we have included the contours between $w_1$ and $w_2$ in Fig. \ref{fig:2d_posterior}, where $w_2$ also enters the positive range, indicating a highly negative correlation with $w_1$, and suggesting a pronounced phantom behavior at higher redshifts. { While both parametrizations contain three free parameters, the PADE form introduces nonlinearities via the denominator, which can lead to a more complex and flexible evolution of the DE EoS. This added flexibility increases the potential for parameter degeneracy, resulting in broader or more elongated contours in the posterior distributions. The broader contours are thus a reflection of the model's internal structure and its ability to accommodate a wider range of dynamical behavior. Finally, we show the contour plot between the matter energy density $\Omega_{m0}$ and $w_1$, which indicates that the mean value of $w_1$ deviates from 0 toward negative values. This deviation suggests a preference for dynamical DE over the cosmological constant. The increasing significance of this trend may point to a mild but growing deviation between $\Lambda$CDM and evolving DE models. The evolution of the EoS parameters as a function of redshift is shown in Fig.~\ref{fig:w_cpl_pade}. For both dataset combinations, the Padé parametrization exhibits a relatively narrower spread in the EoS compared to the CPL model, even at the $2\sigma$ confidence level. This suggests that the functional form of the Padé model inherently limits variation at higher redshifts, thereby influencing the inferred DE behavior.}

\begin{figure}
		\centering
		\begin{minipage}{0.9\linewidth}
			\centering
			\includegraphics[width=\linewidth]{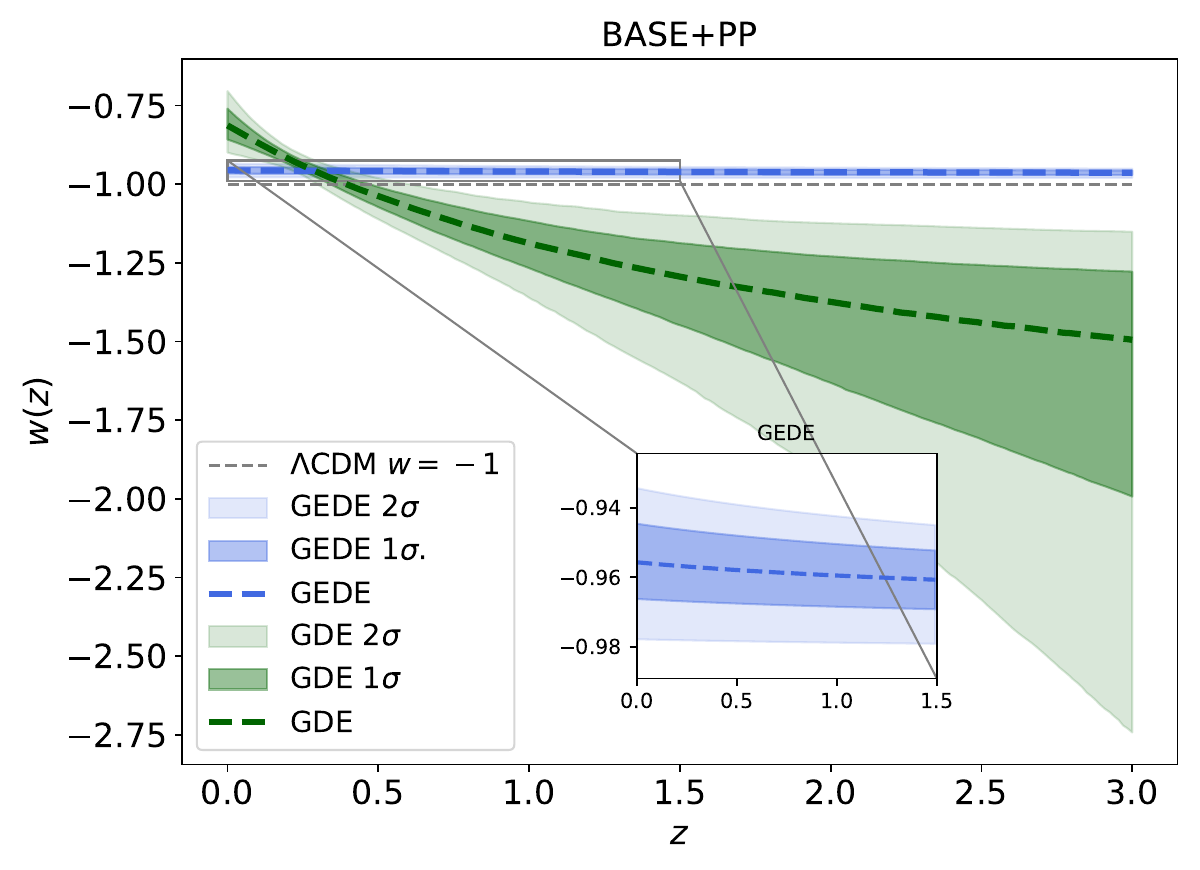}
		\end{minipage}
		\vspace{0.3cm}
		\begin{minipage}{0.9\linewidth}
			\centering
			\includegraphics[width=\linewidth]{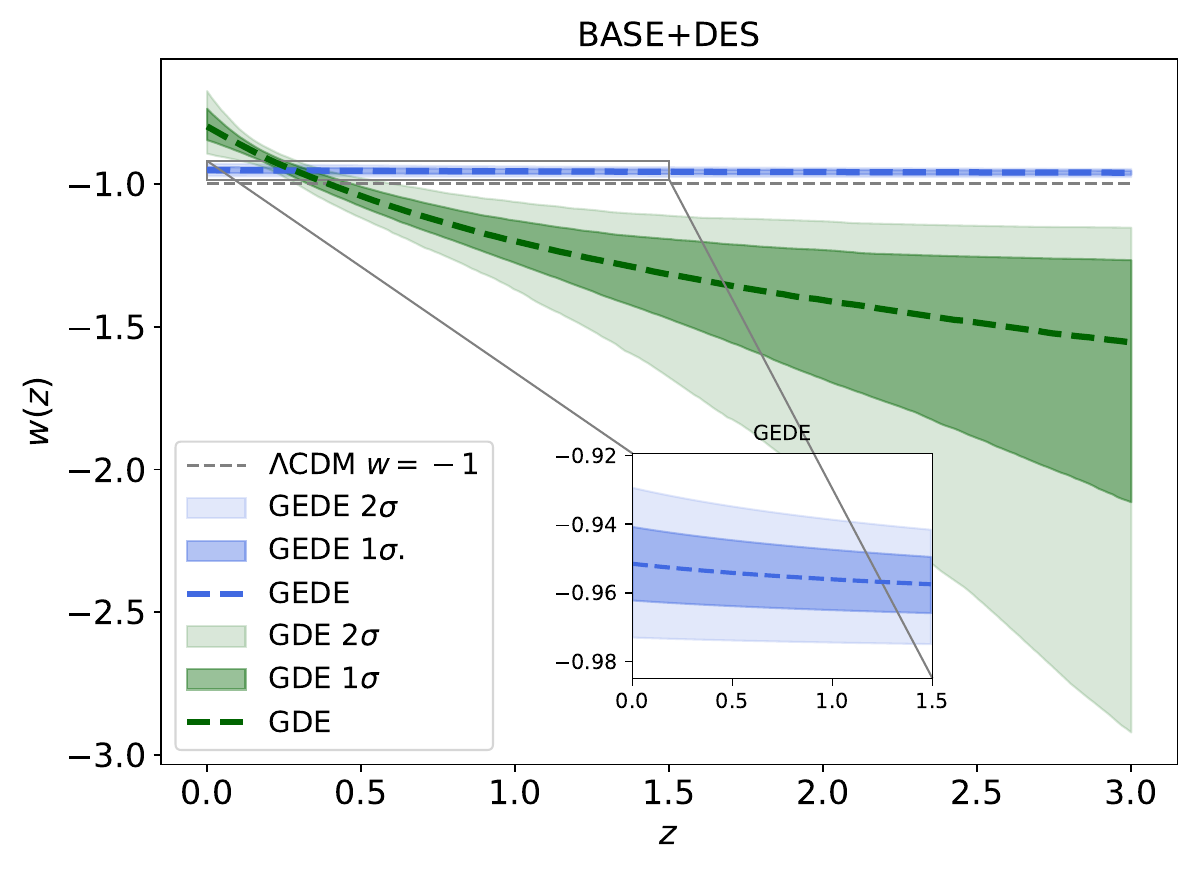}
		\end{minipage}
		\caption{Evolution of $w(z)$ as a function of redshift $z$. The upper panel depicts the behavior of the GEDE and GDE models using the dataset BASE+PP, while the lower panel corresponds to the combination BASE+DESY5. The shaded regions indicate the 68\%~ ($1\sigma$) and 95\%~($2\sigma$) confidence level constraints derived from observational data. The horizontal dashed line in both panels represents the standard $\Lambda$CDM. An inset zoom-in highlights the low-redshift region for enhanced clarity.}
		\label{fig:w_gede_gde}
	\end{figure}

	Figures \ref{fig:gede} and \ref{fig:gde} illustrate the marginalized parameter distribution of the model parameters for the GEDE and GDE models, with the correlation matrix respectively, across all combination of the data sets. Using the MCMC chains, we illustrate the redshift evolution of the EoS of the DE in Fig. \ref{fig:w_gede_gde}. The GEDE model, shown in blue, closely tracks the cosmological constant with \( w \approx -1 \) throughout the redshift range, indicating behavior nearly indistinguishable from the flat \(\Lambda\)CDM model. In contrast, the GDE model demonstrates a significant departure from \( \Lambda \)CDM at higher redshifts, exhibiting a pronounced phantom-like behavior where \( w(z) < -1 \). This deviation becomes more prominent with increasing redshift and is accompanied by broader confidence regions, reflecting the increasing uncertainty in the evolution. Notably, the GDE model accommodates dynamic evolution in the DE sector, allowing for a more flexible phenomenology that may offer insights into early-universe physics or address existing observational tensions. It is important to emphasize that the parameter \( w_0 \) reported in Tab. ~\ref{tab:best_fit} does not directly represent the present-day value of the DE EoS, \( w(z=0) \). To obtain the actual value of \( w \) at redshift \( z=0 \), one must refer to the defining expression given in Eq.~\eqref{eos3}. The redshift evolution and present behavior of \( w(z) \) are illustrated in Fig.~\ref{fig:w_gede_gde}. Despite having the same number of free parameters, GDE and CPL favor stronger deviations into the phantom regime compared to PADE.
	
	It is important to emphasize that the prior ranges for all parameters were chosen uniformly, allowing the models to freely explore both the quintessential (\(w > -1\)) and phantom (\(w < -1\)) regimes without introducing prior-driven bias. As a result, although certain parameterizations may favor phantom behavior at higher redshifts, models with fewer degrees of freedom, such as the GEDE parametrization, can still prefer a quintessential evolution consistent with the data.
	
	{Notably, the GEDE model exhibits predominantly non-phantom behavior and yields AIC values that closely align with those of the $\Lambda$CDM model across both dataset combinations. In contrast, the GDE model more closely resembles those parametrizations that allow a transition from non-phantom to phantom behavior at redshifts $z \gtrsim 0.4$, leading to a better statistical performance relative to $\Lambda$CDM.}

	This raises an important theoretical concern regarding the degree of phantomness, which becomes particularly critical when modeling DE with particle physics and evaluating its implications for cosmic evolution. Although phantom-like behavior (\(w(z)<-1\)) is observationally allowed, it introduces significant challenges within the context of fundamental physics. In scalar field theories, realizing $w<-1$ typically requires a negative kinetic term, resulting in ghost instabilities and violations of unitarity. Such models generally lack a consistent ultraviolet (UV) completion and often exhibit perturbative instabilities, including superluminal propagation and unbounded growth of fluctuations. Additionally, they violate the null energy condition (NEC), raising concerns about causality and the stability of spacetime \cite{Caldwell:2003vq,Ludwick:2017tox}. As a result, any detection of phantom dynamics should be approached with caution and may be more appropriately interpreted as an effective description emerging from modified gravity, non-minimal couplings, or other extensions beyond canonical scalar field frameworks.
	
	These DE parameterizations not only offer flexible descriptions of the cosmic expansion history but also impact the growth of cosmic structures by influencing the timing and dynamics of accelerated expansion, thereby affecting key observables related to matter perturbations. Investigating these effects presents a valuable direction for future studies.\\
		Although the evolution of the EoS up to redshift $z \sim 1$ exhibits minimal differences among the CPL, PADE, and GDE parameterizations, this similarity primarily reflects the current limitations of observational precision at low redshifts. Despite their distinct functional forms, some of the parameterizations can reproduce very similar cosmological observables, such as the expansion rate $H(z)$ and luminosity distance $D_{l}(z)$, for appropriate parameter choices. This degeneracy underscores the challenge of distinguishing between these models with the present data. Our comparison, therefore, aims to highlight that, while these parameterizations provide flexible frameworks to describe DE dynamics, breaking their degeneracy will require more precise measurements (e.g., from future surveys such as Roman or Euclid, see Ref. \cite{Hussain:2025vbo} and references therein) or theoretical guidance to motivate more distinctive model forms.
	
	\section{A Bell-Shaped EOS }
	\label{sec4}
	
	In this section, we propose an alternative parameterization of the DE equation of state (EoS) that we refer to as the \textit{Bell-Shaped} EoS. Its primary purpose is to explore whether a more localized deviation in the EoS around the redshift range where DE is dynamically relevant can be better constrained by the data. The form of the EoS is given by
	\begin{equation}
		w_{\rm de}(z) = w_1 + (w_0 - w_1) \exp\left(-\left(\frac{z-z_t}{\Delta}\right)^2\right) \ ,
	\end{equation}
where $w_0$ and $w_1$ govern the amplitude and asymptotic behavior of the EoS, $z_t$ denotes the redshift of transition, and $\Delta$ sets the width of the bell-shaped feature. Unlike the CPL or Padé forms, the value of $w_{\rm de}(z=0)$ here is not directly equal to $w_0$, but depends on the choice of the rest parameters. 
	
	This functional form is motivated by a key observation: in the standard parameterizations such as CPL, Padé and GDE, the DE EoS often crosses into the phantom regime ($w < -1$) around $z < 0.5$, where observational uncertainties remain large. Furthermore, the EoS in those models tends to become poorly constrained at higher redshifts, leading to wide posteriors or artificial extrapolation. In contrast, the BellDE form localizes the dynamical variation of $w(z)$ and asymptotically approaches a constant value $w_1$ at high redshift, which may better reflect the current state of our observational knowledge.
	
\begin{figure}
		\centering
		\begin{minipage}{0.9\linewidth}
			\centering
			\includegraphics[width=\linewidth]{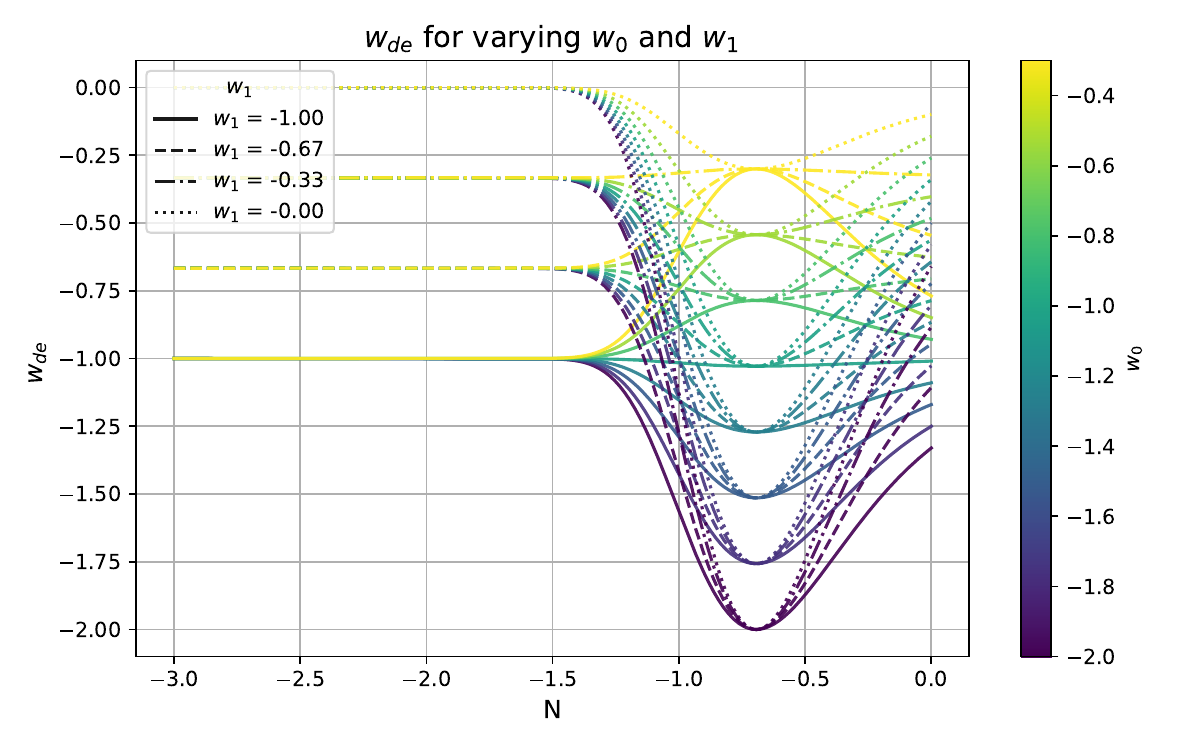}
		\end{minipage}
		\vspace{0.3cm}
		\begin{minipage}{0.9\linewidth}
			\centering
			\includegraphics[width=\linewidth]{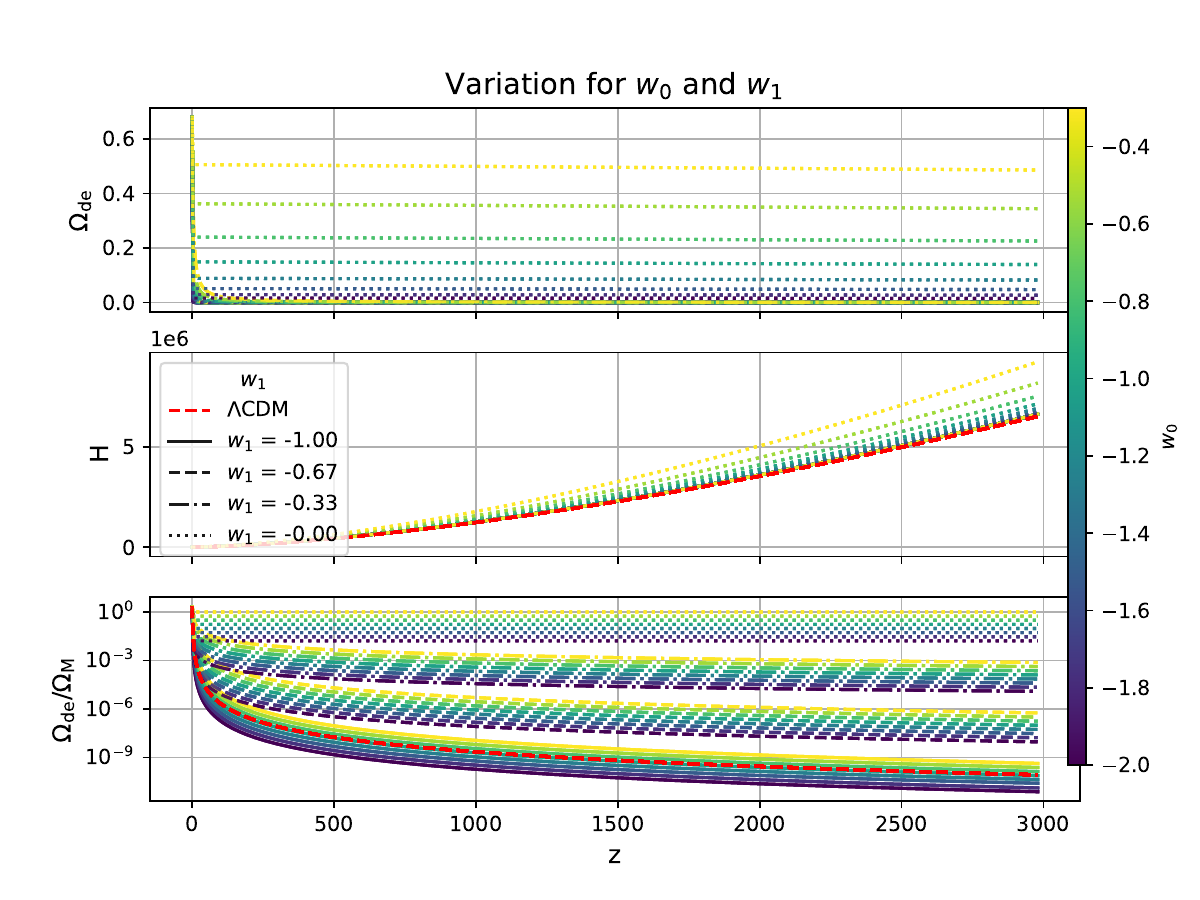}
		\end{minipage}
		\caption{The variation of EoS, energy densities and Hubble parameter of the bell-shaped EoS for $\Delta = 0.95,\ z_t = 1.0, H_0 = 70$ km/s/Mpc. Here, the EoS variation is shown against $N = -\log(1+z).$ }
		\label{fig:eos_new}
	\end{figure}

	While models with three or more EoS parameters often suffer from weak constraints due to data limitations at higher redshift, interpreting phantom-like behavior in such models can be misleading. The BellDE parametrization is specifically designed to offer a smooth, bounded, and flexible evolution of the DE EoS, with particular emphasis on the low-redshift regime where DE exerts its greatest influence. Although its functional form is well-behaved and free from pathological divergences or instabilities, it does not impose strict bounds requiring the EoS to remain exclusively above or below the phantom divide. Unlike standard parameterizations such as CPL or PADE, which typically assume monotonic evolution or a single transition, BellDE accommodates more complex dynamics, including at most double crossing of the phantom divide as can be seen in Fig. \ref{fig:eos_new}. The variation of EoS and the energy density is shown in Fig. \ref{fig:eos_new} with respect to $N = -\log(1+z)$. The EoS asymptotically approaches a constant value governed by the parameter \( w_1 \) at the higher redshift $z>>1$ or $N<<-1$. The lower panel shows the ratio of DE to matter density. Certain parameter combinations—for example, with \(w_1 > -1\)—can lead to non-negligible DE contributions at \(z \gtrsim 1100\), potentially conflicting with constraints from the CMB. In contrast, when \(w_1 \approx -1\), the model closely tracks the behavior of \(\Lambda\)CDM at high redshift, ensuring radiation- and matter-dominated epochs are preserved. We find that combinations with \(w_0 \in [-1.4, -1.0]\) and \(w_1 \sim -1\) yield consistent late- and early-time behavior, making the model viable for structure formation and early universe physics. Hence, BellDE replicates the effective behavior of \(w\)CDM at high redshift, while retaining flexibility at \(z \sim 0.5\)–1, where current data are most sensitive to deviations from \(w = -1\). \\
	The aim of BellDE is not to assert a particular theoretical form of DE but to probe whether allowing localized variations in \(w(z)\), particularly centered around the redshift of matter–DE equality, can yield improved phenomenological fits. Although the parametrization enforces a correlation between low- and high-redshift behavior, this is a generic feature of any fixed functional form. We caution that any precision in the high-redshift EoS should be interpreted carefully, as it largely reflects extrapolation from low-redshift constraints rather than direct observational sensitivity.
	
\begin{figure}
		\centering
		\begin{minipage}{0.9\linewidth}
			\centering
			\includegraphics[width=\linewidth]{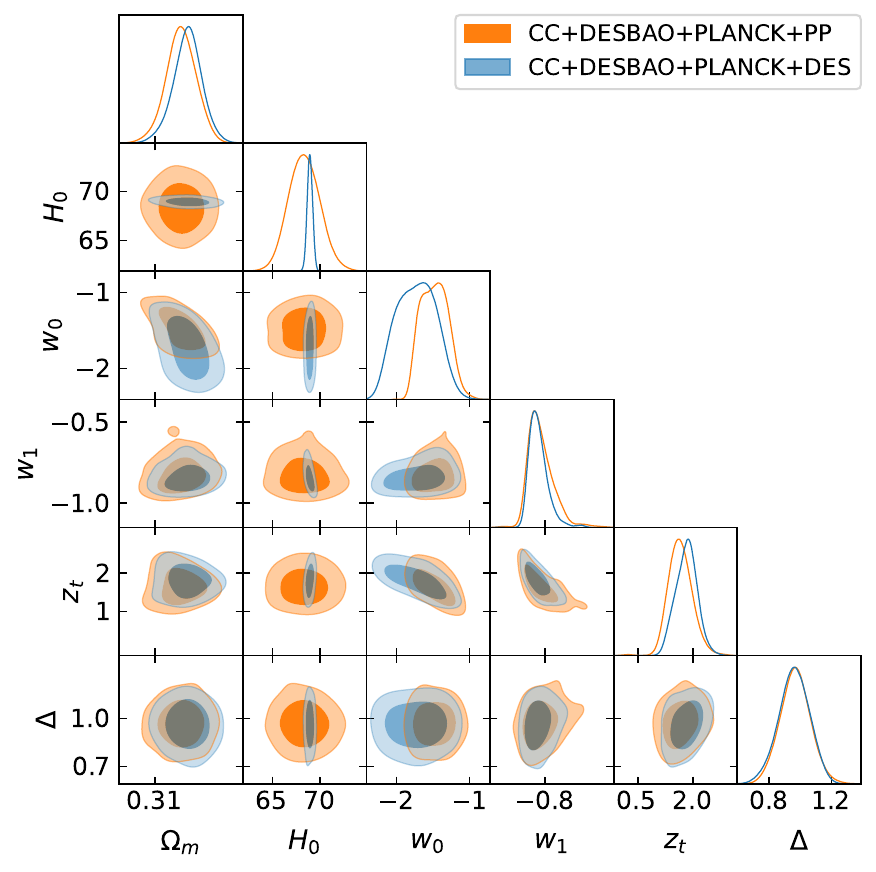}
		\end{minipage}
		\vspace{0.3cm}
		\begin{minipage}{0.9\linewidth}
			\centering
			\includegraphics[width=\linewidth]{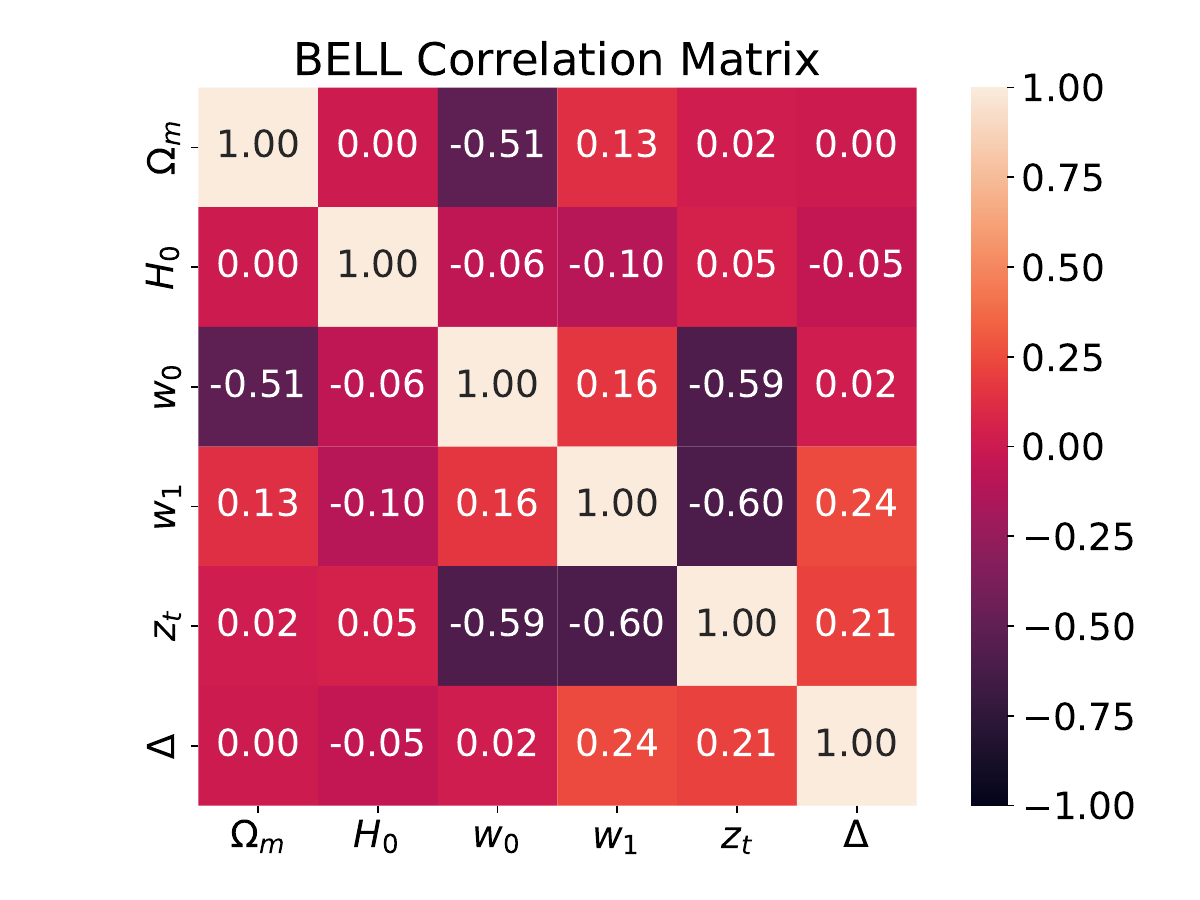}
		\end{minipage}
		\caption{The 1D and 2D marginalized posterior distributions of the model parameters for Bell-Shaped EoS parameterizations, with the nuisance parameter \(M_B\) marginalized over. The correlation matrix is constructed by combining both datasets and marginalizing over \(M_B\).}
		\label{fig:new_bell_corner}
	\end{figure}

	We constrain the model using the observational datasets described earlier, with uniform priors:
	$		w_0 \in [-2.5, 2],\
			w_1 \in [-1.3, 0],\
			z_t \in [0.1, 3],\
			\Delta \sim \mathcal{N}(1.0, 0.1)$
	Here, we treat $\Delta$ with a Gaussian prior motivated by previous fits. The resulting marginalized 1D and 2D posterior distributions, and the correlation matrix, are shown in Fig.~\ref{fig:new_bell_corner}. Strong negative correlations are observed between \((w_0, \Omega_m)\), \((w_0, z_t)\), and \((w_1, z_t)\), while mild positive correlations appear between \((w_1, \Delta)\) and \((z_t, \Delta)\).

\begin{figure}
		\centering
		\begin{minipage}{0.9\linewidth}
			\centering
			\includegraphics[width=\linewidth]{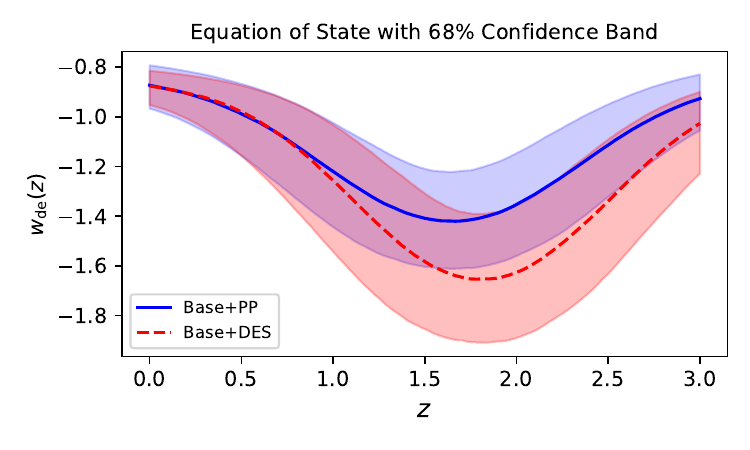}
		\end{minipage}
		\vspace{0.3cm}
		\begin{minipage}{0.9\linewidth}
			\centering
			\includegraphics[width=\linewidth]{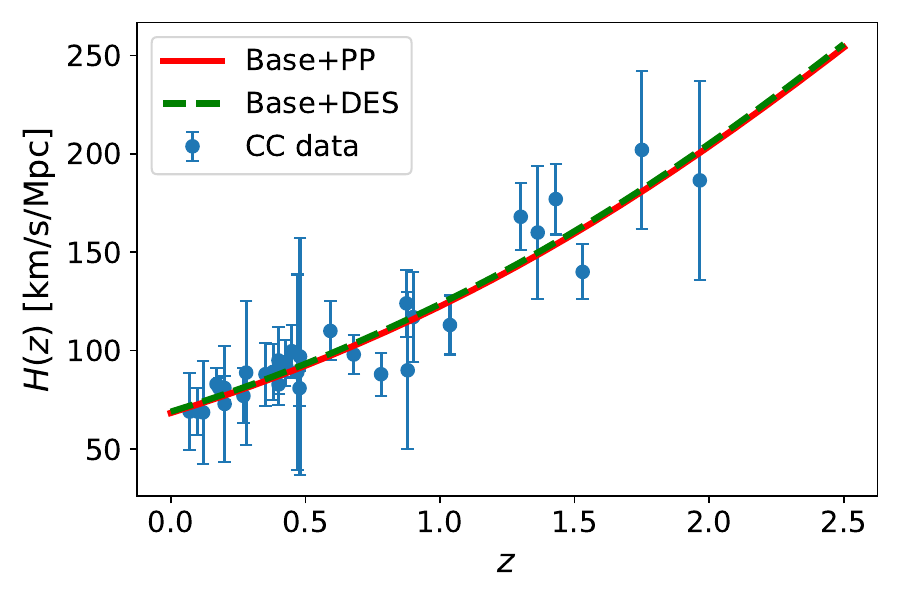}
		\end{minipage}
		\vspace{0.3cm}
		\begin{minipage}{0.9\linewidth}
			\centering
			\includegraphics[width=\linewidth]{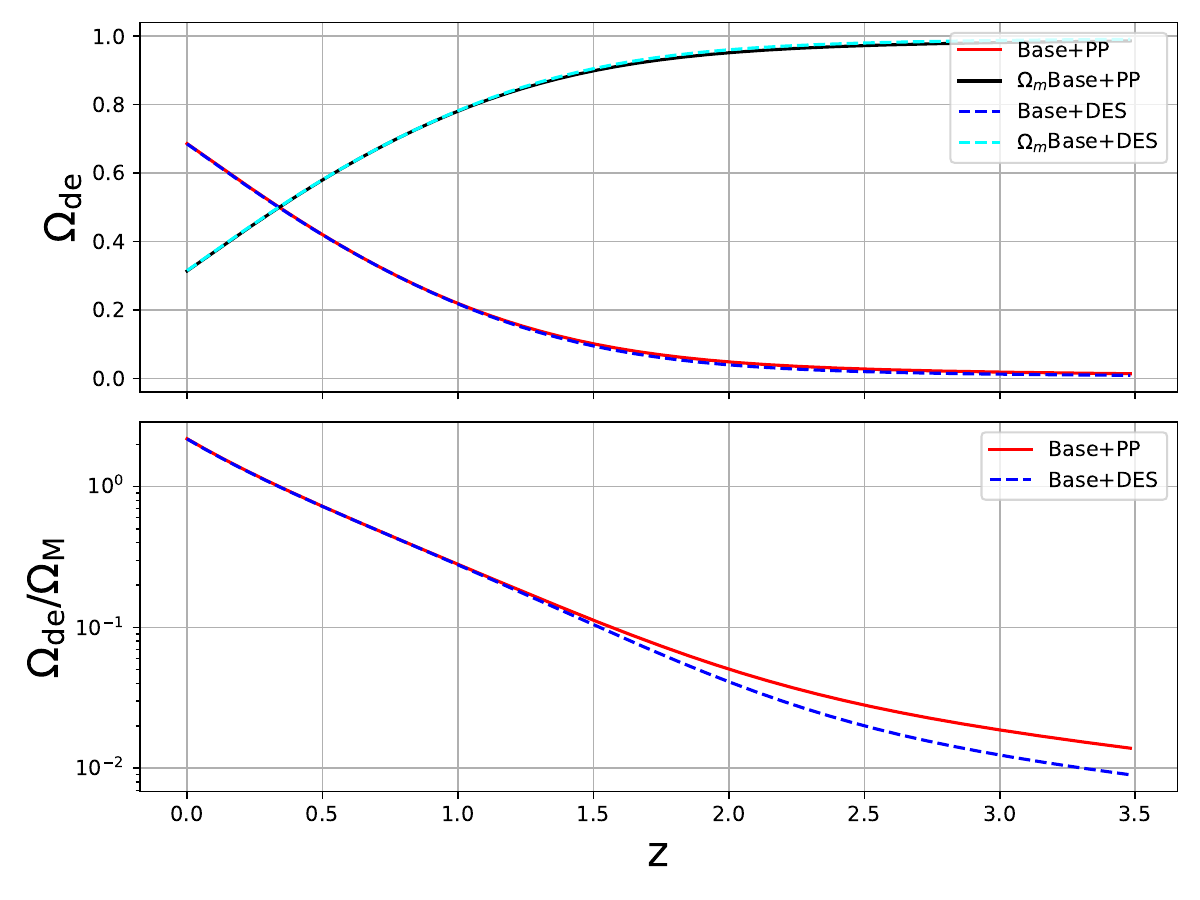}
		\end{minipage}
		\caption{$1\sigma$ level error propagation of the bell-shaped EoS and the corresponding cosmological parameters evaluated at their best-fit values.}
		\label{fig:error_bell_eos}
	\end{figure}

	The mean values of the model parameters are summarized in Tab.~\ref{tab:best_fit}. The evolution of the EoS, along with the associated uncertainty derived from the mean parameter values, is shown in Fig.~\ref{fig:error_bell_eos}. Despite having the same number of free parameters as CPL or PADE, BellDE yields tighter constraints on certain parameters, such as \(w_0\), due to its localized structure.
	
	The BellDE model yields \(w(z=0) = -0.862\) (BASE+PP) and \(w(z=0) = -0.866\) (BASE+DES), values that are close to the cosmological constant and consistent with current bounds. The reduced \(\chi^2\) values are nearly unity, and the AIC values are comparable to those of CPL, indicating a good fit without overfitting. In particular, the BellDE model provides competitive statistical performance while offering greater flexibility in the redshift regime \(z \sim 0.5 - 1\), where DE becomes observationally significant.

	\section{Summary and conclusions} 
	\label{sec5}
	
	\begin{figure}
		\centering
		\begin{minipage}{0.9\linewidth}
			\centering
			\includegraphics[width=\linewidth]{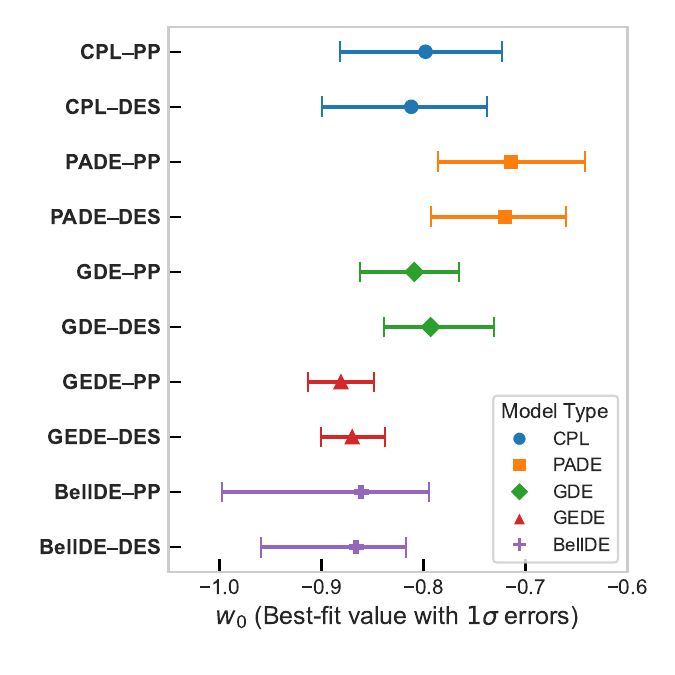}
		\end{minipage}
		\vspace{0.3cm}
		\begin{minipage}{0.9\linewidth}
			\centering
			\includegraphics[width=\linewidth]{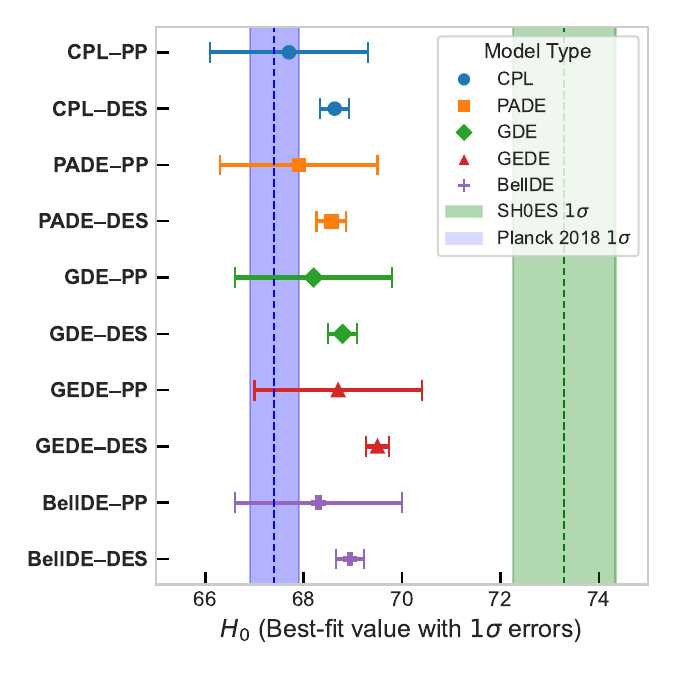}
		\end{minipage}
		\caption{Whisker plot of $w_{0}$ and $H_0$ of the considered models with the 68\% CL constraints, for datasets used in our analysis. For example, CPL-PP and CPL-DES refer to the CPL model constrained with BASE+PP and BASE+DES, respectively; similar notation applies to the other models.}
		\label{fig:er}
	\end{figure}
	
	In this work, we investigate a variety of DE EoS parameterizations, including extensions beyond the conventional CPL form, to explore potential deviations from the cosmological constant value \( w_\Lambda = -1 \). Motivated by the 2025 DESI data release, we have updated our analysis of dynamical DE reconstructions by incorporating the latest DESI DR2 BAO measurements alongside data from the Pantheon+ and DESY5 supernova compilations. This comprehensive approach enables us to trace the temporal evolution of the DE EoS with improved precision. {Our findings show that, while the general shape of the reconstructed \( w(z) \) remains consistent with recent DESI results, there is a notable increase in the statistical evidence supporting dynamical DE. Several dynamical DE models achieve lower AIC values relative to the flat \(\Lambda\)CDM model, reflecting a preference for an evolving DE EoS. This preference is particularly enhanced by the inclusion of DESI BAO data, which play a critical role in breaking parameter degeneracies. Consequently, cross-validation of results using multiple DE parameterizations is imperative to ensure the robustness and reliability of these conclusions.}
	
	We initiated our study using a polynomial form of the Taylor series expansion of $w(a)$ up to the second order, which we refer to as the extended CPL. Through our analysis, we found that the EoS for DE at high redshifts is likely to differ significantly from the standard $\Lambda$CDM model (i.e., $w(z) < -1$), suggesting that $w(a) > -1$ in the present day. To further investigate the potential interpretations of this observed deviation's physical origin, we considered three additional model categories.  { The parameterizations considered in this work, including CPL, PADE, GDE, GEDE, and BellDE, are not derived from a fundamental physical theory of DE. Rather, they serve as phenomenological frameworks aimed at capturing a broad spectrum of possible behaviors of the DE EoS \( w(z) \) in a model-independent manner, particularly in the absence of a well-established underlying theory.
		Their differing mathematical structures allow them to probe distinct aspects of \( w(z) \), such as smooth transitions, phantom crossings, and variable evolutionary features. In particular, the BellDE form is capable of accommodating multiple transitions between phantom and non-phantom regimes, which is not feasible in simpler two-parameter models like CPL. Although all models tend to converge within the redshift range most tightly constrained by current data, their extrapolated behavior beyond this region and the statistical preference for or against certain trends can provide meaningful insights into the types of DE evolution that are observationally viable.}
	
	 {As we observed, when DESY5 is included (refer to Fig.~\ref{fig:er}), there is evidence for a higher $H_{0}$ compared to when PP is present. In most cases, particularly for the CPL and PADE models, we observed that the uncertainties on $w_0$ and $H_0$ are only marginally reduced, or even slightly increased, when PP is used instead of DES. This suggests that the addition of PP does not significantly improve the constraining power over DES, at least within the context of the combined dataset with CC and DESI BAO. Such as, the error bars on $w_0$ for PADE–PP and PADE–DES are nearly identical, while the difference in $H_0$ error bars is modest. }
	
	Another category of DE models we examine is the generalized emergent class, where DE has had little to no influence throughout most of cosmic history and only begins to emerge in more recent times. It is important to note that while this model can replicate the emergence of DE, it is constrained by its inability to cross $w = -1$, as is anticipated even with late-time datasets. Specifically, any apparent crossing in the $w(a)$ parameterization could be misleading in efforts to align with observables, prompting the question of whether $w(a)$ actually crosses $-1$ or if the observed behavior is merely a result of the parameterization artefact. After completing this work, the authors became aware of a related study by Scherer et al. \citep{Scherer:2025esj}, in which a transient EoS for DE was proposed. Their findings support the exploration of transient DE models, which are favored by various datasets and exhibit notable deviation from the $\Lambda$CDM model.
	
	Finally, we conducted an assessment of the novel EoS we proposed, designated ``BellDE". This bell-shaped EoS inherently facilitates the possibility of reversing the transition, enabling a return to conditions where $w > -1$, thereby restoring quintessential behavior at higher redshifts. While the number of free parameters associated with this model is comparable to those found in previously discussed EoS models, the uncertainties pertaining to the parameters of the bell-shaped model are significantly smaller. This observation underscores its enhanced capacity for constraint. Furthermore, unlike prior models that frequently display substantial phantom behavior at elevated redshifts, the bell-shaped EoS effectively mitigates such behavior in the early universe, approaching a non-phantom value asymptotically. 
	 
		In particular, the BellDE form permits multiple transitions between phantom and non-phantom regimes—a feature not achievable in simpler two or three-parameter models such as CPL. While all parameterizations tend to converge within the redshift range where current data are most constraining, their behavior outside this region, along with the statistical preference for specific trends, can offer valuable insights into the forms of DE evolution that are observationally permitted or favored. This study highlights that several DE parameterizations consistently favor an evolving EoS when DESI BAO data are included. These trends provide a clear understanding of the model-dependent correlations between low and high-redshift behavior in interpreting the constraints on \(w(z)\).

	The forthcoming supernova observations from the Nancy Grace Roman Space Telescope \citep{Hounsell:2023xds}, Euclid, and LSST are expected to extend the accessible redshift range, thereby tightening constraints on the DE EoS $w_{0}$. These advancements position the next decade as potentially transformative for cosmology, with the prospect of determining whether a paradigm shift is necessary in our understanding of the Universe. As the precision and volume of cosmological data continue to increase, tensions between different measurement approaches have become more pronounced. The field thus stands at a pivotal moment, requiring a careful reassessment of both observational strategies and the theoretical framework underpinning modern cosmology.

\subsection*{Analysis at effective redshift}

Previous studies have examined the implications of DESI observations for the $\Lambda$CDM framework, emphasizing that if an extended model such as $w_0w_a$CDM provides a statistically better fit, this preference should be reflected in the redshift evolution of the $\Lambda$CDM parameters themselves \citep{Colgain:2021pmf,Colgain:2024mtg,Colgain:2024xqj,Colgain:2025nzf}. In such analyses, consistency across multiple cosmological probes is crucial for distinguishing genuine physical effects from statistical fluctuations or observational systematics, particularly those that may arise in the LRG sample at $z_{\mathrm{eff}} = 0.51$. Following this motivation, we also analyze the DESI BAO DR2 measurements at their corresponding effective redshifts within our extended models to test the stability of model parameters and assess potential departures from $\Lambda$CDM.

\begin{table*}
	\centering
	\renewcommand{\arraystretch}{1.15}
	\setlength{\tabcolsep}{4pt}
	\begin{tabular}{c
			cc|cc|cc|cc|cc|cc}
		\hline\hline
		\multirow{2}{*}{$z_{\mathrm{eff}}$} 
		& \multicolumn{2}{c}{CPL} 
		& \multicolumn{2}{c}{PADE} 
		& \multicolumn{2}{c}{BellDE} 
		& \multicolumn{2}{c}{$\Lambda$CDM} &&&&\\
		\cmidrule(lr){2-3}\cmidrule(lr){4-5}\cmidrule(lr){6-7}\cmidrule(lr){8-9}\cmidrule(lr){10-11}\cmidrule(lr){12-13}
		& $100 H_0 r_d$ & $\Omega_m$
		& $100 H_0 r_d$ & $\Omega_m$
		& $100 H_0 r_d$ & $\Omega_m$
		& $100 H_0 r_d$ & $\Omega_m$ &&&&\\
		\hline
		0.51 & $93.21^{+5.0}_{-7.2}$ & $0.394^{+0.13}_{-0.074}$ & $95.95^{+6.5}_{-7.4}$ & $0.315^{+0.15}_{-0.099}$ & $94.64^{+3.7}_{-4.30}$   & $0.307^{+0.14}_{-0.088}$ &  $97.62 \pm {2.5}$   & $0.372^{+0.047}_{-0.061}$ &&&&\\
		0.71 & $102.90^{+8.3}_{-12}$ & $0.367^{+0.085}_{-0.074}$ & $102.90 \pm 8.9$ & $0.297^{+0.12}_{-0.088}$ & $99.57^{+4.2}_{-5.7}$  & $0.292^{+0.13}_{-0.060}$ &  $100.4 \pm {2.5}$   & $0.338^{+0.039}_{-0.045}$ &&&&\\
		0.93 & $106.80^{+9.4}_{-13}$ & $0.303^{+0.078}_{-0.070}$ & $107.0 \pm 9.8$ & $0.249^{+0.093}_{-0.07}$ &  $103.40^{+5.1}_{-7.0}$  & $0.261^{+0.077}_{-0.037}$ &  $102.9 \pm {2.1}$   & $0.281 \pm {0.025}$ &&&&\\
		1.32 & $101.90^{+8.8}_{-13}$ & $0.304 \pm 0.088$ & $101.50 \pm 20$ & $0.30 \pm 0.12$ &  $100.60^{+6.3}_{-7.3}$   & $0.257^{+0.082}_{-0.044}$ & $102.8 \pm {3.1}$   & $0.286^{+0.027}_{-0.033}$ &&&&\\
		1.48 & $101.70^{+9.5}_{-13}$ & $0.336 \pm 0.089$ & $103.0 \pm 11$ & $0.288 \pm 0.092$ &  $98.04^{+6.3}_{-8.6}$   & $0.302^{+0.088}_{-0.066}$& $98.06 \pm {4.9}$   & $0.327^{+0.044}_{-0.058}$ &&&&\\
		2.33 & $98.15^{+8.2}_{-11}$ & $0.320 \pm 0.087$ & $102.60 \pm 9.8$ & $0.273 \pm 0.083$ &  $98.43^{+6.8}_{-7.9}$ & $0.300^{+0.056}_{-0.046}$ &  $100.6 \pm {3.7}$   & $0.305^{+0.026}_{-0.032}$ &&&&\\
		\hline \hline
		\multirow{2}{*}{$z_{\mathrm{eff}}$} 
		& \multicolumn{2}{c}{GDE} 
		& \multicolumn{2}{c}{GEDE}  &&&&&&&& \\
		\cmidrule(lr){2-3}\cmidrule(lr){4-5}
		& $100 H_0 r_d$ & $\Omega_m$
		& $100 H_0 r_d$ & $\Omega_m$  &&&&&&&& \\
		\hline
		0.51 & $93.18^{+5.2}_{-8.0}$ & $0.420^{+0.13}_{-0.083}$ & $94.52^{+3.1}_{-3.6}$ & $0.323^{+0.10}_{-0.061}$ &&&&&&&& \\
		0.71 &  $103.0^{+8.5}_{-13}$ & $0.383^{+0.10}_{-0.089}$  & $97.82 \pm 3.70$ &  $0.308^{+0.073}_{-0.047}$  &&&&&&&& \\
		0.93  & $106.40^{+9.6}_{-16}$ & $0.326^{+0.082}_{-0.094}$ & $100.50^{+4.2}_{-3.4}$  & $0.257^{+0.051}_{-0.022}$ &  &&&&&&& \\
		1.32 & $102.30^{+9.6}_{-14}$ & $0.317^{+0.088}_{-0.098}$ & $98.76 \pm 5.2$ & $ 0.271^{+0.053}_{-0.030}$ &  &&&&&&& \\
		1.48 &  $101.40^{+11}_{-15}$ & $0.34 \pm 0.11$ &  $95.92 \pm 5.6$ & $0.319 \pm 0.066$  &&&&&&&&  \\
		2.33 &  $99.23^{+8.7}_{-11}$ & $0.320 \pm 0.087$ & $97.82 \pm 5.4$ & $0.319^{+0.031}_{-0.041}$  &&&&&&&& \\
		\hline \hline
	\end{tabular}
	\caption{Comparison of $H_0 r_d$ and $\Omega_m$ at each effective redshift 
		between $\Lambda$CDM and alternative dark energy models using DESI BAO DR2 + CC data, where for each BAO redshift we use the correlation matrix between $D_M/r_d$ and $D_H/r_d$.}
	\label{tab:DESI_BAO}
\end{table*}

\begin{figure*}
    \centering
    \includegraphics[width=0.8\linewidth]{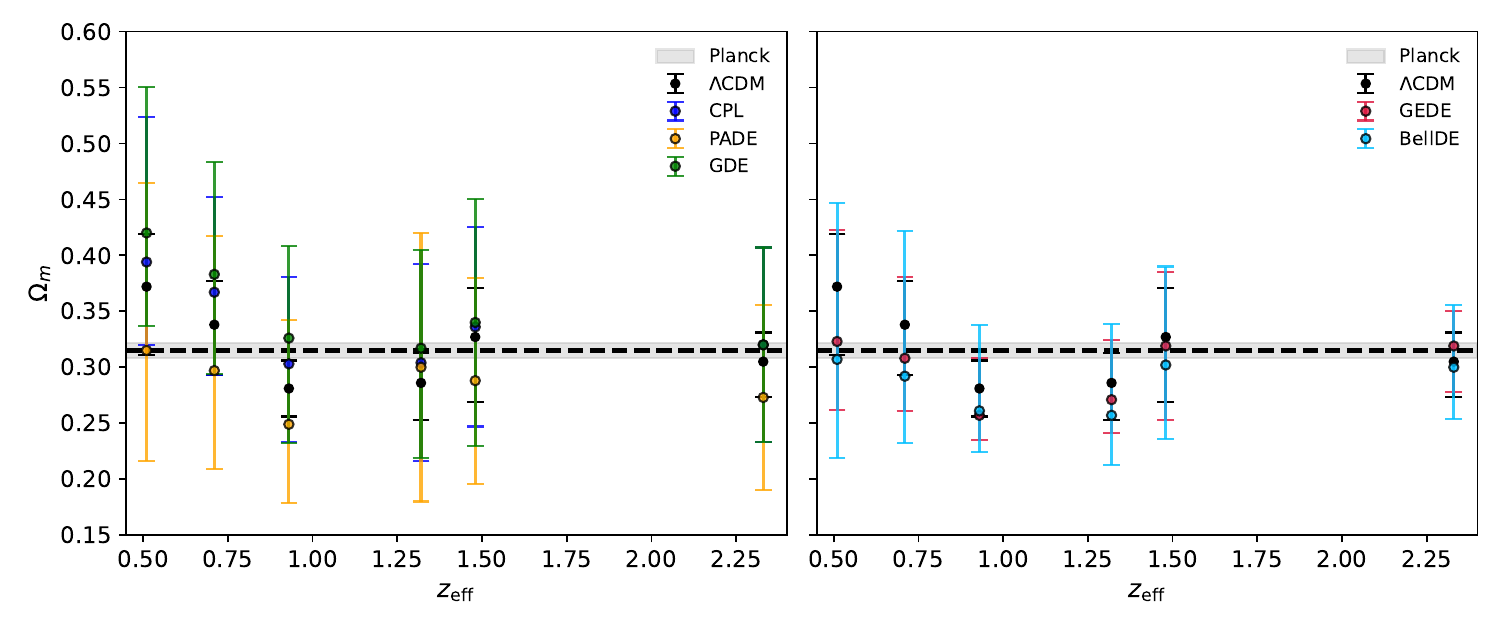}
    \caption{68\% credible intervals for the matter density parameter $\Omega_m$ as a function of effective redshift $z_{\mathrm{eff}}$ for different dark energy models using DESI BAO DR2 + CC data. Panel~(a) compares CPL, PADE, and GDE with $\Lambda$CDM, while panel~(b) shows GEDE and BellDE. The dashed line and gray band indicate the Planck reference ($\Omega_m = 0.315 \pm 0.007$).}
    \label{Omega}
\end{figure*}

We adopt a standard broad uniform prior of $\Omega_m \in [0,1]$ and $H_0 r_d \in [0,10^{6}]$. It is important to emphasize that no model parameter was fixed; all parameters were allowed to vary freely during the MCMC analysis. The motivation behind this choice is that any deviation from the $\Lambda$CDM model should manifest as systematic variations in $\Omega_m$ across different redshifts when interpreted within the framework of dynamical dark energy models. Following the methodology outlined in Ref.~\cite{Colgain:2025fct}, we perform independent analyses at each effective redshift listed in Table~\ref{tab:DESI_BAO}, thereby allowing us to probe the potential redshift evolution of the model parameters.\footnote{In addition to the DESI BAO DR2 data, we also include the cosmic chronometer (CC) data to obtain an effective estimate of $\Omega_m$. While the inclusion of external datasets such as Hubble measurements shifts the central value of $\Omega_m$, it does not alleviate the underlying statistical discrepancy reported in Table~1 of Ref.~\cite{Colgain:2024xqj}. For a fair comparison, we therefore present the results using the same dataset as that considered for the $\Lambda$CDM case.} Consequently, we find a mild indication of deviation from $\Lambda$CDM, which can be interpreted as evidence for a dynamical dark energy component.

The results of the analysis at each DESI BAO redshift are summarized in Table~\ref{tab:DESI_BAO}, where the fitted parameters $\Omega_{\rm M}$ and \(100\,H_0 r_d\) are reported. For the $\Lambda$CDM model, the matter density at the effective redshift $z_{\rm eff}=0.51$ is found to be comparatively higher than at other redshifts. This evolution, though within the 1–2$\sigma$ level, it may hint at small internal inconsistencies if interpreted strictly under a static $\Lambda$CDM background. A similar trend is also observed for alternative models such as CPL and GDE. In contrast, the PADE model yields a slightly lower value of $\Omega_{\rm M}$ at $z_{\rm eff}=0.93$, while remaining consistent across the other redshift bins. Notably, the BellDE and GEDE models exhibit a stable behavior, maintaining a consistent $\Omega_{\rm M}$ across all redshift bins, thereby indicating better agreement with the data compared to the other models. This analysis reveals that certain dynamical dark energy models perform relatively better than the $\Lambda$CDM model. At the same time, the results indicate that not all forms of dynamical dark energy are compatible with the data, thereby providing a useful framework to distinguish among different model classes.

Overall, the differences among the models, although statistically modest, are physically informative. The spread in $H_0 r_d$ remains mild; however, the non-monotonic behavior of $\Omega_m(z_{\mathrm{eff}})$ underscores that the DESI+CC data are sensitive to departures from a purely constant-$w$ cosmology, as illustrated in Fig.~\ref{Omega}. These findings further motivate testing dynamical dark energy frameworks beyond $\Lambda$CDM. The apparent redshift dependence of the best-fit parameters may reflect either residual systematics in the low-$z$ BAO sample or genuine signatures of evolving cosmic acceleration.

The broader context of this study lies in the persistent $\Lambda$CDM tensions, most notably the discrepancies in $H_0$ and $S_8$, which, if physical in origin, suggest that the $\Lambda$CDM framework may fail due to cosmological parameters evolving with effective redshift. Supporting this interpretation, several independent analyses have reported a systematic decrease in the inferred $H_0$ and a corresponding increase in $\Omega_m$ with increasing effective redshift across multiple cosmological probes~\citep{Colgain:2024mtg}. If confirmed, these results would point to a breakdown of the $\Lambda$CDM description at the background level in the late Universe. It would be valuable to further explore this direction using full-shape clustering data in future work, as such a detailed investigation lies beyond the scope of the present analysis. A similar strategy has been adopted in recent studies~\citep{Colgain:2025fct}, which offer a useful framework for extending the present results.

\begin{table}
\centering
\begin{tabular}{lcccc}
\hline\hline
Model & BASE+PP & BASE+DES  &&\\
\hline
CPL &  $2.8\sigma$ & $4.1\sigma$ &&\\
PADE    &  $2.7\sigma$ & $4.2\sigma$ &&\\
GDE & $2.5\sigma$ & $3.9\sigma$  &&\\
BellDE & $2.4\sigma$ & $3.8\sigma$&& \\
GEDE & $2.2\sigma$ & $3.3\sigma$ && \\
\hline
\end{tabular}
\caption{Results reproduced from Table \ref{tab:best_fit}, showing the inferred $H_0$ tensions with SH0ES. }
\label{tab:DESI_Tensions}
\end{table}

\subsection*{$H_0$ tension}

Table~\ref{tab:DESI_Tensions} summarizes the statistical significance of the Hubble tension between the SH0ES measurement and the $H_0$ values inferred from various dark energy parameterizations. The results are based on the best-fit values reported in Table~\ref{tab:best_fit} for two dataset combinations, namely BASE+PP and BASE+DES. Across all models, the inclusion of DES data leads to a higher level of tension with SH0ES, consistent with the tighter constraints and slightly lower $H_0$ estimates obtained in those fits. Among the models considered, CPL and PADE yield the largest discrepancies ($\sim4.1$–$4.2\sigma$), in agreement with the findings of Refs.~\citep{Colgain:2025nzf,Lee:2022cyh,Alestas:2020mvb}. In contrast, the GEDE and BellDE models exhibit comparatively lower tensions ($\sim2$–$3\sigma$), suggesting that their dynamical features allow a modest alleviation of the $H_0$ tension while maintaining consistency with the combined cosmological datasets. This trend supports the view that frameworks incorporating time-dependent dark energy dynamics may absorb part of the Hubble tension at the background level, underscoring the importance of extending such analyses with full-shape clustering and high-$z$ BAO data in future work.

	\section*{Acknowledgements}
	S.H. acknowledges the support of the National Natural Science Foundation of China under Grant No. W2433018 and No. 11675143, and the National Key Research and Development Program of China under Grant No. 2020YFC2201503. S.A. acknowledges the Japan Society for the Promotion of Science (JSPS) for providing a postdoctoral fellowship during 2024-2026 (JSPS ID No.: P24318). This work of SA is supported by the JSPS KAKENHI grant (Number: 24KF0229). A.W. is partially supported by the US NSF grant: PHY-2308845. B.R. is partially supported as a member of the Roman Supernova Project Infrastructure Team under NASA contract 80NSSC24M0023. We would like to thank the editor and the anonymous reviewer for their comments and suggestions, which significantly helped us improve our work.
    
	\section*{Data Availability}
Data sharing is not applicable to this article as no datasets were generated during the current study.
	
\bibliographystyle{mnras}
	\bibliography{Ref}

\end{document}